\documentclass[aps,prd,amsmath,floats,floatfix,
twocolumn, superscriptaddress,nofootinbib,showpacs]{revtex4-1}

\usepackage{amsfonts,amsmath,units,wasysym,epsfig,graphicx,verbatim,color,subfigure,graphicx}
\usepackage{amsmath}
\usepackage{amssymb}
\usepackage{amsfonts}
\usepackage{bm}
\usepackage{color}
\usepackage[normalem]{ulem}
\usepackage{xcolor}

\def\be{\begin{equation}}
\def\ee{\end{equation}}
\def\bea{\begin{eqnarray}}
\def\eea{\end{eqnarray}}

\begin{document}


\newcommand{\rhat}{\hat{r}}
\newcommand{\iotahat}{\hat{\iota}}
\newcommand{\phihat}{\hat{\phi}}
\newcommand{\h}{\mathfrak{h}}
\newcommand{\vek}[1]{\boldsymbol{#1}}
\newcommand{\IUCAA}{\affiliation{Inter-University Centre for Astronomy and Astrophysics, Post Bag 4, Ganeshkhind, Pune 411 007, India}}
\newcommand{\WSU}{\affiliation{Department of Physics \& Astronomy, Washington State University, 1245 Webster, Pullman, WA 99164-2814, U.S.A}}
\newcommand{\UBC}{\affiliation{UBC}}
\newcommand{\SU}{\affiliation{Syracuse}}
\newcommand{\PENN}{\affiliation{Department of Physics, The Pennsylvania State University, University Park, PA 16802, USA}}
\newcommand{\UNH}{\affiliation {Department of Physics, University of New Hampshire, 9 Library Way, Durham NH 03824, USA}}
\newcommand{\LBL}{\affiliation{Lawrence Berkeley National Laboratory, 1 Cyclotron Rd, Berkeley, CA 94720, USA} }
\newcommand{\KEK}{\affiliation{Theory Center, Institute of Particle and Nuclear Studies, KEK, Tsukuba 305-0801, Japan}}
\newcommand{\SOKEN}{\affiliation{Department of Particle and Nuclear Physics, the Graduate University for Advanced Studies (Sokendai), Tsukuba 305-0801, Japan}}
\newcommand{\ITHEMS}{\affiliation{Interdisciplinary Theoretical and Mathematical Sciences Program (iTHEMS), RIKEN, Wako, Saitama 351-0198, Japan}}
\newcommand{\KYOTO}{\affiliation{Center for Gravitational Physics, Yukawa Institute for Theoretical Physics, Kyoto University, Kyoto 606-8502, Japan}}
\newcommand{\TAPIR}{\affiliation{TAPIR, Walter Burke Institute for Theoretical Physics,
    California Institute of Technology, MC 350-17, Pasadena, CA 91125, USA}}
\newcommand{\UWM}{\affiliation{Department of Physics, University of Wisconsin-Milwaukee, Milwaukee, Wisconsin 53201, USA}}
\newcommand{\Cornell}{\affiliation{Cornell Center for Astrophysics and Planetary Science, Cornell University, Ithaca, New York, 14853, USA}}
\newcommand{\CITA}{\affiliation{Canadian Institute for Theoretical
    Astrophysics, University of Toronto, Toronto, Ontario M5S 3H8, Canada}}
\newcommand{\UofT}{\affiliation{Department of Astronomy \& Astrophysics,
    University of Toronto, Toronto, Ontario, M5S 3H5, Canada}}
\newcommand{\AEI}{\affiliation{Max Planck Institute for Gravitational Physics (Albert Einstein Institute), D-14476 Potsdam-Golm, Germany}} %
\newcommand{\SYRACUSE}{\affiliation{Department  of  Physics,  Syracuse  University,  Syracuse,  NY  13244,  USA}}

\title{Systematic effects from black hole-neutron star waveform model uncertainties on the neutron star equation of state}


\author{Kabir Chakravarti}\IUCAA
\author{Anuradha Gupta}\PENN\IUCAA
\author{Sukanta Bose}\IUCAA\WSU
\author{Matthew D. Duez}\WSU
\author{Jesus Caro}\WSU
\author{Wyatt Brege}\WSU
\author{Francois Foucart}\UNH\LBL
\author{Shaon Ghosh}\UWM
\author{Koutarou Kyutoku}\KEK\SOKEN\ITHEMS\KYOTO
\author{Benjamin D. Lackey}\AEI
\author{Masaru Shibata}\AEI\KYOTO
\author{Daniel A. Hemberger}\TAPIR
\author{Lawrence E. Kidder}\Cornell
\author{Harald P. Pfeiffer}\AEI\CITA
\author{Mark A. Scheel}\TAPIR

\date{\today}


\begin{abstract}

We identify various contributors of systematic effects in the measurement of the neutron star (NS) tidal deformability and quantify their magnitude for several types of neutron star - black hole (NSBH) binaries. Gravitational waves from NSBH mergers
contain information about the components' masses and spins as well as the NS equation of state.  Extracting this information requires comparison of the signal in noisy detector data with theoretical templates derived from some combination of post-Newtonian (PN) approximants, effective one-body (EOB) models and 
numerical relativity (NR) simulations. 
The accuracy of these templates is limited by errors in the NR simulations, by the approximate nature of the PN/EOB waveforms, and by the hybridization procedure used to combine them.

In this paper, we estimate the impact of these errors by constructing and comparing a set of PN-NR hybrid waveforms, for the first time with NR waveforms from two different codes, namely, SpEC and SACRA, for such systems. We then attempt to recover the parameters of the binary using {\it two} non-precessing template approximants.
As expected, these errors 
have negligible effect on detectability. Mass and spin estimates are moderately affected by systematic errors for near equal-mass binaries, while the recovered masses can be inaccurate at higher mass ratios. Large uncertainties are also found in the tidal deformability $\Lambda$, due to differences in PN base models used in hybridization, numerical relativity NR errors, and inherent limitations 
of the hybridization method.  We find that systematic errors are too large for tidal effects to be accurately characterized for any realistic NS equation of state model.  We conclude that NSBH waveform models must be significantly improved if they are to be useful for the extraction of NS equation of state information or even for distinguishing NSBH systems from binary black holes.  

\end{abstract}
\preprint{[LIGO-PXXXXXXX]}

\maketitle



\section{Introduction}
\label{intro}

Compact object binaries are of paramount importance in gravitational wave (GW) physics because their signals are loud, have a significant rate of occurrence, and are amenable to theoretical modelling. It is perhaps only a matter of time before Neutron Star-Black Hole (NSBH) binaries 
are discovered \cite{Rates2010,Rates2016}.
The phenomenological modeling of such systems is thus an important problem today. From the point of view of tidal effects, an NSBH binary constitutes a simplified version of the binary neutron star (BNS) system, as it involves only one tidal parameter. Consequently, any inference drawn on the NS equation of state (EOS) from their GW signal would be cleaner than that from BNSs.

A few phenomenological (Phenom) models of NSBH binary waveforms already exist. Post-Newtonian (PN) and numerical relativity (NR) waveforms were first combined to make complete NSBH waveform families by Lackey~{\it et al.}~\cite{Lackey_2014}.  They created two template families, based on different analytical 
waveform models. One used the 
aligned-spin IMRPhenomC binary black hole (BBH) model~\cite{Santamaria_etal}, while the other used a time-domain effective one-body (EOB) model~\cite{Taracchini_etal_2012}.  The merger phase and amplitude were then modified by analytic correction functions calibrated to 134 NR simulations produced by the
SACRA code~\cite{SACRA}.  We will henceforth call these waveforms the ``LEA'' model.  As an extension of LEA, Pannarale~{\it et al.}~\cite{Pannarale2015a} introduced a more detailed frequency-domain model for the waveform amplitude by distinguishing systems with various degrees of expected tidal disruption.
Finally, Kumar {\it et al.}~\cite{Prayush_2016b}, as part of a study of systematic errors in mass and spin estimates due to tidal effects, produced an enhanced version of the LEA templates, using SEOBNRv2 \cite{Taracchini_etal_2014} as the underlying BBH model, with the same tidal corrections as LEA. We call the resulting templates the ``LEA+'' model. 

In this paper, we identify key systematic errors that can affect the measurement of the NS tidal deformability parameter, $\Lambda$, in possible NSBH detections.
This is the first time that results impacting NSBH parameter estimation
are obtained by comparing waveforms based on
simulations carried out with two different NR codes, namely, SpEC          
\cite{Kidder:2000yq} and SACRA \cite{SACRA}. The paper is organized as follows: In Sec.~\ref{syst} we summarize the different sources of systematic errors in the tidal deformability estimation. We also describe the NR late-inspiral-merger-ringdown waveforms and the procedure for constructing hybrid waveforms by combining them with PN inspiral cycles. 
In Sec.~\ref{results}, we present the results of parameter estimation for these hybrid waveforms using the LEA and LEA+ models, before concluding in Sec.~\ref{end}.

\section{waveform systematics}
\label{syst} 


The characterization of NSBH binaries involves the estimation of a NS tidal parameter, apart from the masses and spins of the binary components. Here we will take the dimensionless tidal deformability ~\cite{Lattimer:2012nd} to represent the former.
We follow the LEA convention and denote this parameter by $\Lambda \equiv (2/3)\,k_2\,(c^2\,R_{\rm NS}/G\,M_{\rm NS})^5$, where $k_2$,
$R_{\rm NS}$ and $M_{\rm NS}$ are the second Love number, radius and mass, respectively, of the neutron star. There are multiple sources of systematics that will affect the error budget for $\Lambda$. At the outset they can be grouped into the following broad categories.

\begin{enumerate}

\item {\bf NR modeling:} To begin with, there will be systematics due to errors in the NR waveforms,
namely, the effects of finite resolution and of the extrapolation of waveforms to infinity.
In the limit of infinite resolution and perfect initial data, all evolution methods are equivalent, but with the introduction of a finite grid (or finite extraction radius), the equivalence is broken and so numerical errors become dependent on the process chosen to evolve the field equations. 
There will also be errors due to imperfect initial data leading to, for example, non-zero residual eccentricities or `junk' GW radiation.

\item {\bf Choice of EOS model:} 
The choice of EOS model may also influence parameter estimation. To first order, $\Lambda$ is the best measured matter property of the NS from the pre-merger phase of the GW signal from an NSBH. Yet, the space of EOSs is not one dimensional, and two different stars with the same mass and $\Lambda$ can be constructed from two different EOS models, e.g., (somewhat simplistically) a single polytrope and a piecewise polytrope. In this sense, the choice of the EOS model used by NR can distort the GW waveform computed 
because the true EOS presumably does not fall perfectly within the EOS family used in the NR survey, albeit
possibly only during the last few orbits of inspiral and during/after merger.

\item {\bf Choice of PN waveform:} When constructing hybrid waveforms, the choice of the  
PN approximant used is also a potential source of error. 

\item {\bf The PN-NR bridge:} The method for combining the PN and NR cycles to produce complete waveforms in the detector band -- also termed as ``hybridization'' -- can be a source of systematic error. This is true even if the technique itself is perfect, in that it produces an accurate complete waveform if the PN and NR parts themselves are accurate. This is because
as long as even one set of cycles (PN or NR) is erroneous or the number of overlapping PN and NR cycles is too few, the hybridization procedure can create a waveform that is inaccurate even at frequencies where the original PN or NR parts were in themselves accurate.

\item {\bf Sampling the ambiguity function in $\Lambda$:} 
Accurate parameter estimation depends on the knowledge and application of accurate waveform models. Estimating the value of a signal parameter then involves cross-correlating normalized templates based on such waveforms with the GW data containing a signal. As detailed here, the rate at which this cross-correlation (or, more precisely, the {\em match}, as defined below) changes with $\delta \Lambda \equiv \Lambda_{\rm template} - \Lambda_{\rm signal}$ is slow: When the signal itself is a unit-norm ``neighboring" template, the match drops by a few tenths of a percent even when $\delta \Lambda$ is as high as several tens to a few hundreds. This is just another way of stating that the ambiguity function~\cite{Dhurandhar:2017aan} in $\Lambda$ is diffuse or not sharply peaked. For strong signals, this spread in $\Lambda$ will reduce, and its estimation will be more {\em precise}. It does not impact estimation accuracy by itself. However, we show here that even for reasonably accurate templates if the sampling rate of the data (and the templates) is not considerably higher than 4096~Hz, the match can have multiple closely-spaced local maxima, which can cause parameter estimation algorithms to miss the global maximum. This effect can make $\Lambda$ estimation inaccurate.

\item {\bf The mass-ratio effect:} The mass ratio of the binary (i.e., $q \equiv M_{\rm BH}/M_{\rm NS}$, where $M_{\rm BH}$ and $M_{\rm NS}$ are the black hole and neutron star masses, respectively) inversely dictates the magnitude of the tidal effect in its signal. With increasing $q$, tidal effects on the waveform are reduced. Systematic errors like the ones we discussed start to dominate over the physical tidal effects, and this means that it is harder to estimate the tidal parameter accurately. Note that this is in addition to the {\em statistical} effect whereby a smaller value of a parameter will incur a larger spread in its measured value. The high mass ratios also tend to make NR simulations harder.

\end{enumerate}

The first five systematic errors can be reduced through improved modeling or better parameter estimation algorithms and their implementation in data in the future. The last effect is related to actual physical processes, and tells us where modeling errors will impact parameter estimation the most (i.e., at high mass ratios).

\subsection{Systematics from Numerics}

Since 
tidal effects 
are expected to be small, 
{\bf numerical errors} can significantly impact measurements of the tidal deformability in NSBH binaries.
In this paper, we employ waveforms from two of the major NR codes used to model NSBH mergers:  SACRA~\cite{SACRA} and SpEC~\cite{Kidder:2000yq}. These two codes were written independently of each other and use different formulations of general relativity, grid structures, and numerical methods.  The SACRA waveforms utilized here are from the same set that was used to calibrate LEA.  
Some of the SpEC waveforms come from simulations recently used to study the impact of dynamical tides on NSBH waveforms~\cite{Hinderer_etal_2016}, while others are presented for the first time here (see below).
Both SpEC and SACRA simulations have drastically improved in phase accuracy in recent years with the introduction of sophisticated techniques such as higher-order hydrodynamics~\cite{Radice:2013xpa} and constraint propagation~\cite{Bernuzzi:2009ex}. Since we use waveforms made over a period of several years, 
numerical errors vary significantly from system to system.

There are several possible sources of NR error. One source is residual eccentricity in the initial data. SpEC simulations use an iterative approach~\cite{Pfeiffer:2007yz} to reduce the initial eccentricity, resulting in residual eccentricities $e\sim 0.0005-0.003$ for the simulations presented here.  
The SACRA simulations employed to calibrate LEA, on the other hand, used quasi-circular initial data, with typical residual eccentricities $e\sim 0.01$.
Another source of error is the extrapolation to future null infinity. In SpEC, this error is estimated by comparing 
extrapolation methods of different orders~\cite{Boyle:2009vi}, and is found to be small compared to other numerical errors (phase errors of $\sim 0.01\,{\rm rad}$)~\cite{Foucart:2012vn}.
For the SACRA simulations used in LEA, gravitational waves were extracted at a fixed radius of $\sim 1000\,\mathrm{km}$ instead.
We note that the eccentricity reduction and extrapolation to null infinity are both performed in BNS simulations by SACRA~\cite{Kyutoku:2014yba}, and that updated NSBH simulations are ongoing.

At the current accuracy of numerical simulations, however, the largest source of error is truncation error. The magnitude of that error can be estimated by comparing simulations at different resolutions, and in some cases by extrapolating results to infinite resolution. The most recent SpEC simulations (cf. Cases 1 and 2 in Tables~\ref{tab-resolution} and \ref{tab-gesolution}) accumulate phase difference of about 0.1 - 0.3 radian at merger over $20-30$ waveform cycles.  Older simulations have errors several times larger, for both SpEC and SACRA.

These errors must be compared to the expected impact of finite size effects on the phase of the GW signal.  This is a very strong function of the mass ratio of the binary. Finite size effects are negligible for $q \gtrsim 6$, non-spinning binary. For $q=6$, the result of BBH simulations falls within the error bars of NSBH simulations~\cite{Foucart:2013psa}. On the other hand, even our most conservative error estimates predict that finite size effects are resolved with $\sim 20\%$ relative errors at the time of merger for the lowest mass ratios considered here ($q=1-2$, non-spinning binaries~\cite{Hinderer_etal_2016}). Finally, for high mass-ratio binaries with rapidly spinning black holes, the rapid falloff of the GW amplitude due to the disruption of the NS is the strongest finite-size effect on the waveform, and is well captured by simulations.

Analytical waveform templates may also have errors induced by their calibration to a set of
NR simulations using a restricted family of EOS, i.e., EOS with a fixed functional form and a set of freely specifiable parameters that can capture some but not all of the properties of real NSs. Errors from this effect are expected to be small because, to leading order, finite size effects only depend on the tidal deformability of NSs, and the same range of $\Lambda$ can be covered with many EOS families.  The SACRA waveforms used for LEA assume a two-component piecewise polytropic EOS, i.e., an equation of state in which the pressure $P$ is related to the baryon density $\rho$ by
\begin{equation}
P=\begin{cases}
\begin{array}{c}
\kappa_0 \rho^{\Gamma_0} \quad  \text{if}\,\, \rho<\rho_t \\
\kappa_1 \rho^{\Gamma_1} \quad  \text{if} \,\,\rho>\rho_t\,,\\
\end{array}\end{cases}
\end{equation}
where $\kappa_0=3.5966\times 10^{13}$ in cgs units and $\Gamma_0 = 1.3569$.  This EOS family has two free variables, which are taken to be $\Gamma_1$, the high-density polytropic exponent, and $P_1$, the pressure at a fiducial density $\rho_\mathrm{fidu} =10^{14.7}$g cm${}^{-3}$. It has been suggested in Ref.~\cite{Lattimer:2000nx} that the pressure at this numerical value of $\rho_\mathrm{fidu}$ is 
correlated with the NS radius. The value of $\rho_t$ is determined from these quantities by requiring the continuity of the pressure at the interface as $\rho_t = ( \kappa_0 / \kappa_1 )^{1/( \Gamma_1 - \Gamma_0 )}$ with $\kappa_1 = P_1 / \rho_\mathrm{fidu}^{\Gamma_1}$. In this paper, we will use NR waveforms generated with this EOS family as well as those generated using a single-component polytropic EOS:  $P=\kappa\rho^{\Gamma}$, with $\Gamma=2$. 

\subsection{Systematic Errors from templates} 

In Sec.~\ref{results}, we compare PN-NR hybrid waveforms to the LEA and LEA+ models.
Our results are impacted by 
systematic errors in these models. For example, there exists known residual discrepancy between the LEA waveforms and the SACRA simulations used in their construction.  According to Lackey~{\it et al.}~\cite{Lackey_2014}, systematic errors in the analytical Phenom waveform due to this fitting error are $|\Delta\Lambda^{1/5}|/\sigma_{\Lambda^{1/5}}\sim 0.1$ (worse for $q>4$), where $\sigma_{\Lambda^{1/5}}$, the aLIGO statistical error for a source at 100 Mpc, is between 0.5 and 1 (see Fig. 15 of \cite{Lackey_2014}).  So
$|\Delta\Lambda|/\Lambda \approx
5 |\Delta\Lambda^{1/5}|/\Lambda^{1/5}
\approx 5-12\%$,
perhaps twice as big for $q$ between 4 and 5.  Systematic errors in the analytical EOB waveforms due to their discrepancy from their simulation input are about twice as big as those in the analytical Phenom waveforms~\cite{Lackey_2014}. 
Further, as an example of inbuilt template systematic errors, there are known instances for LEA where the phase of the calibrating hybrid waveform is known to have significant amplitude mismatch with the analytical template (Figs. 21 and 22 of Ref.~\cite{Lackey_2014}), all of which add to the systematic error budget.

Systematic errors in the LEA and LEA+ templates are also impacted by the choice of underlying point-particle (BBH) waveform, as the models are constructed by adding tidal terms to BBH templates. Naturally, the performance of the tidal templates depends on the quality of the BBH model.
This is the key difference between LEA and LEA+, and we will see that this non-trivially affects $\Lambda$ estimates when we perform comparisons between LEA and LEA+.

\subsection{PN Model and Hybridization Systematics} 
\label{subsec:hybrid}

As we noted earlier, 
the choice of PN model and the method used to construct the full inspiral-merger-ringdown hybrid waveforms are themselves potential sources of systematic errors.
In this study, the hybrid waveforms are 
constructed using the procedure outlined in Ref.~\cite{Ajith2008}, unless mentioned otherwise. We use the 3.5PN phase corrected SpinTaylorT4 PN approximant, with 1.5PN amplitude correction whereas tidal corrections are taken up to 6PN order.
SpinTaylorT4 is one of the popular inspiral models and matches remarkably well with NR in the case of equal mass, non-spinning binaries~\cite{Boyle:2009vi}, whereas for non-equal mass binaries, it is of comparable accuracy as other Taylor-approximants~\cite{MacDonald:2011ne}.  Therefore, we use the SpinTaylorT4 model for the inspiral part of the hybrid waveforms. The systematics due to the use of different PN approximants in constructing hybrids for non-precessing NSBH systems will be studied in a future publication.

The hybridization is performed in the time domain.
We minimize the integrated absolute squared difference between PN and NR waveforms in a time window where we have NR data and where 
the PN approximation is expected to be valid:
\begin{equation}
\label{eq:delta}
\delta=\int_0^T |h_{PN}(t)-a*h_{NR}(t,\vec{\mu})|^2dt \,.
\end{equation}
Here $[0,T]$ is
the aforementioned overlap region of PN and NR waveforms in the time domain. We choose $t=0$ to be an instantaneous signal frequency that is immediately after the point when the junk radiation in the NR signal was emitted. On the other hand, $T$ is chosen more adaptively by allowing for as many cycles of the PN waveform into the minimization domain as possible without coming closer than a few cycles of the last stable circular orbit (LSCO). The minimization of $\delta$ is carried out over an amplitude scaling factor $a$ and the extrinsic parameters $\vec{\mu}$ of the NR waveforms, namely, the initial phase $\phi_0$ and the initial time of arrival $t_0$.
Once the minimizing values (i.e., $a_{\rm min}$ and $\vec{\mu}_{\rm min} = \{ \phi_{0 {\rm min}}, t_{0 {\rm min}}\}$) are obtained, we make use of a linear interpolation 
joining function to produce the hybrid waveforms.  The waveforms so produced can be expressed as:
\begin{equation}
\label{eq:hybrid}
h_{\text{hyb}}(t,\vec{\mu})
= h_{\text{PN}}(t)[1-\tau(t)] \\
+a_{\text{min}}h_{\text{NR}}(t,{\vec{\mu}}_{\text{min}})\tau(t) \,,
\end{equation} 
where $\tau(t)$ is the linear interpolation function, given by
\begin{equation}
\label{eq:hybridtau}
\tau (t)=\begin{cases}
\begin{array}{c}
0 \quad  \text{if}\quad t\leq t_{0\text{min}} \\
\frac{t-t_{0\text{min}}}{t_{\text{max}}-t_{0\text{min}}} \quad\text{if}\quad t_{0\text{min}}<t\leq  t_{\text{max}}\\
1  \quad  \text{if}\quad t_{\text{max}}<t \,. \\
\end{array}\end{cases}
 \\ 
\end{equation}
Here, $t_{\rm max}$ is the time up to which we take the PN waveform to be valid; it is typically somewhat before the LSCO. 
Such an interpolation requires that NR cycles are available not just for the merger and ringdown phases, but also for at least several cycles 
of the late-inspiral phase at separations larger than the LSCO.
This is the method used to produce the hybrids shown in Fig.~\ref{Fig:hybrids}.

It is important to note here that the ultimate test of the quality of these hybrid waveforms is how well the hybrid matches a NR waveform with the same binary parameters that covers the complete band of the signal in Eq.~(\ref{eq:hybrid}). Phenomenological waveform models are approximations of such waveforms that are usually deduced by requiring high fitting factors against the hybrids. Such a criterion, however, does not guarantee that those waveform models will also estimate signal parameters highly accurately. Note that phenomenological models are calibrated to imperfect hybrid waveforms, and thus are themselves imperfect - even if they match perfectly the waveform they are calibrated against. Details on errors that can result from the hybridization procedure have been examined in Ref.~\cite{MacDonald:2011ne}. Even though it analyzed only binary black hole systems, its findings have relevance for NSBH systems studied here, with the main  difference being that the latter include an additional parameter, in the form of the NS tidal deformability $\Lambda$.

It is also worth mentioning that 
this hybridization method fails for our shortest NR waveforms, as the matching interval $[0,T]$ does not contain enough GW cycles. When that is the case, we
adaptively modify the function $\tau$ in Eq.~(\ref{eq:hybrid}) to a step function. 
The hybrid waveform is then
\begin{equation}
\label{eq:adhybrid}
h_{\text{hyb}}(t,\vec{\mu})
= h_{\text{PN}}(t)[1-\tau(t)] \\
+h_{\text{NR}}(t,{\vec{\mu}}_{\text{min}})\tau(t) \,,
\end{equation} 
where
\begin{equation}
\tau (t)=\begin{cases}
\begin{array}{c}
0 \quad  \text{if}\quad t\leq t_{0\text{min}} \\
1  \quad  \text{if}\quad t_{0\text{min}}<t \,. \\
\end{array}\end{cases}
 \\ 
\end{equation}
$t_{0\text{min}}$ is defined as in Eqs.~(\ref{eq:hybrid}) and (\ref{eq:hybridtau}).

Henceforth, we will refer to this method as the ``step" method of hybridization.

\begin{table}[h]
\begin{tabular}{| p{1.5cm} | p{1.4cm} |  p{1cm} | p{1cm}| p{1cm} | p{1cm}| c|}
    \hline
\multicolumn{2}{|c|}{Injected Waveform} & \multicolumn{4}{|c|}{Best Matched Parameters} & FF \\
 \hline
 Resolution & Code & $M_{\rm BH}$ & $M_{\rm NS}$ & $\chi_{\rm BH}$ & $\Lambda$ & $\%$ \\
 \hline
\multicolumn{7}{|l|}{Case 1. $q=1.5$; $\chi_{\rm BH}=0$; $M_{\rm NS}=1.4$; EOS $\Gamma=2$; $\Lambda=791$}\\
\hline
low & SpEC & 2.06  & 1.42 & 0.001 & 579.0 & 99.80 \\
medium & SpEC & 2.06  & 1.41 & -0.005 & 546.3 & 99.80 \\
high & SpEC & 2.10  & 1.39 & -0.006 & 551.3 & 99.77 \\
\hline
\hline
\multicolumn{7}{|l|}{Case 2. $q=2$; $\chi_{\rm BH}=0$; $M_{\rm NS}=1.4$; EOS $\Gamma=2$; $\Lambda=791$}\\
\hline
low & SpEC & 2.76  & 1.41 & -0.008 & -785.0 & 99.72 \\
medium & SpEC & 2.79  & 1.40 & -0.002 & -659.1 & 99.71 \\
high & SpEC & 2.71  & 1.43 & -0.021 & -742.9 & 99.75 \\
\hline
\hline
\multicolumn{7}{|l|}{Case 3. $q=2$; $\chi_{\rm BH}=0.75$; $M_{\rm NS}=1.2$; EOS 2H; $\Lambda=4382$}\\
\hline
medium & SpEC & 2.35  & 1.21 & 0.813 & 4097 & 98.94\\
high & SACRA & 2.53  & 1.22 & 0.816 & 3977 & 99.86\\
\hline
\hline
\multicolumn{7}{|l|}{Case 4. $q=3$; $\chi_{\rm BH}=0$; $M_{\rm NS}=1.4$; EOS $\Gamma=2$; $\Lambda=620$}\\
\hline
low & SpEC & 4.00  & 1.46 & -0.031 & -107.7 & 98.97 \\
medium & SpEC & 4.00  & 1.46 & -0.027 & -103.8 & 98.95 \\
high & SpEC & 4.00  & 1.46 & -0.034 & -2288 & 99.49 \\
\hline
\hline
\multicolumn{7}{|l|}{Case 5. $q=3$; $\chi_{\rm BH}=0$; $M_{\rm NS}=1.35$; EOS H; $\Lambda=607$}\\
\hline
 medium & SpEC* & 3.86  & 1.40 & -0.038 & -6358 & 99.41 \\
 high & SpEC* & 3.99  & 1.37 & -0.000 & 594.1 & 97.70\\
 low & SACRA & 3.79  & 1.43 & -0.042 &-426.7  & 99.05 \\
medium & SACRA & 3.86  & 1.40 & -0.036 & -3750 & 99.61 \\
high & SACRA & 3.86  & 1.40 & -0.029 & -891 & 99.09 \\
\hline
\hline
\multicolumn{7}{|l|}{Case 6. $q=5$; $\chi_{\rm BH}=0.75$; $M_{\rm NS}=1.35$; EOS 2H; $\Lambda=2324$}\\
\hline
low & SACRA* & 5.70  & 1.53 & 0.859 & 2367 & 97.46 \\
medium & SACRA* & 5.92  & 1.49 & 0.861 &16784 & 97.31 \\ 
high & SACRA* & 5.82  & 1.51 & 0.849 & 121.5 & 97.58 \\
\hline
\hline
\multicolumn{7}{|l|}{Case 7. $q=5$; $\chi_{\rm BH}=0.5$; $M_{\rm NS}=1.4$; EOS $\Gamma=2$; $\Lambda=791$}\\
\hline
low & SpEC & 5.28  & 1.73 & 0.475 & 1008 & 99.27 \\
medium & SpEC & 5.30  & 1.72 & 0.476 & 997 & 99.28 \\
\hline
\end{tabular}
\caption{
The parameter estimates and fitting factors for a set of hybridized NSBH waveforms, constructed from NR late-inspiral and merger cycles at different resolutions from SpEC and SACRA, when match-filtered with LEA templates in aLIGO noise PSD. 
For example, Case 1 shows three resolutions of SpEC waveforms with $q=1.5$ and $\Lambda=791$ recovered against LEA templates. Asterisk-marked cases were  hybridized using the step method.}
\label{tab-resolution}
\end{table}

\begin{table}[h]
\begin{tabular}{| p{1.5cm} | p{1.4cm} |  p{1cm} | p{1cm}| p{1cm} | p{1cm}| c|}
    \hline
\multicolumn{2}{|c|}{Injected Waveform} & \multicolumn{4}{|c|}{Best Matched Parameters} & FF \\
 \hline
 Resolution & Code & $M_{\rm BH}$ & $M_{\rm NS}$ & $\chi_{\rm BH}$ & $\Lambda$ & $\%$ \\
 \hline
\multicolumn{7}{|l|}{Case 1. $q=1.5$; $\chi_{\rm BH}=0$; $M_{\rm NS}=1.4$; EOS $\Gamma=2$; $\Lambda=791$}\\
\hline
low & SpEC & 2.08  & 1.41 & 0.000 & 587.6 & 99.40 \\
medium & SpEC & 2.07  & 1.41 & -0.000 & 723.6 & 99.39 \\
high & SpEC & 2.06  & 1.45 & -0.015 & 373.5 & 99.41\\
\hline
\hline
\multicolumn{7}{|l|}{Case 2. $q=2$; $\chi_{\rm BH}=0$; $M_{\rm NS}=1.4$; EOS $\Gamma=2$; $\Lambda=791$}\\
\hline
low & SpEC & 2.75  & 1.42 & -0.004 & 0.0 & 99.88 \\
medium & SpEC & 2.78  & 1.40 & -0.003 & 78.7 & 99.88 \\
high & SpEC & 2.78  & 1.40 & -0.003 & 44.9 & 99.87 \\
\hline
\hline
\multicolumn{7}{|l|}{Case 3. $q=2$; $\chi_{\rm BH}=0.75$; $M_{\rm NS}=1.2$; EOS 2H; $\Lambda=4382$}\\
\hline
medium & SpEC & 2.47  & 1.16 & 0.760 & 4960 & 98.45 \\
high & SACRA & 2.38  & 1.20 & 0.770 & 4322 & 99.87\\
\hline
\hline
\multicolumn{7}{|l|}{Case 4. $q=3$; $\chi_{\rm BH}=0$; $M_{\rm NS}=1.4$; EOS $\Gamma=2$; $\Lambda=620$}\\
\hline
low & SpEC & 3.82  & 1.52 & -0.046 & 663.6 & 98.97 \\
medium & SpEC & 3.76  & 1.54 & -0.059 & 796.2 & 98.95 \\
high & SpEC & 3.81  & 1.52 & -0.042 & 580.4 & 99.49 \\
\hline
\hline
\multicolumn{7}{|l|}{Case 5. $q=3$; $\chi_{\rm BH}=0$; $M_{\rm NS}=1.35$; EOS H; $\Lambda=607$}\\
\hline
 medium & SpEC* & 3.62  & 1.48 & -0.056 & 278.9 & 99.48 \\
 high & SpEC* & 3.63  & 1.48 & -0.055 & 272.3 & 99.49\\
 low & SACRA & 3.72  & 1.45 & -0.041 &375.8 & 99.49 \\
medium & SACRA & 3.61  & 1.48 & -0.063 & 366.3 & 99.65 \\
high & SACRA & 3.72  & 1.45 & -0.040 & 456.9 & 99.54 \\
\hline
\hline
\multicolumn{7}{|l|}{Case 6. $q=5$; $\chi_{\rm BH}=0.75$; $M_{\rm NS}=1.35$; EOS 2H; $\Lambda=2324$}\\
\hline
low & SACRA* & 6.30  & 1.42 & 0.846 & 2298 & 97.67 \\
medium & SACRA* & 5.79  & 1.52 & 0.846 &64.1 & 97.93 \\ 
high & SACRA* & 6.25  & 1.43 & 0.845 & 1896 & 97.71 \\
\hline
\hline
\multicolumn{7}{|l|}{Case 7. $q=5$; $\chi_{\rm BH}=0.5$; $M_{\rm NS}=1.4$; EOS $\Gamma=2$; $\Lambda=791$}\\
\hline
low & SpEC & 6.04  & 1.56 & 0.502 & 475.6 & 99.41 \\
medium & SpEC & 6.11  & 1.54 & 0.509 & 795.2 & 99.35 \\
\hline
\end{tabular}
\caption{Same as Table~\ref{tab-resolution}, but now employing LEA+ templates instead of LEA.}
\label{tab-gesolution}
\end{table}

In passing, we note that the regular method of hybridization is more likely to fail if there are large values of $q$ or $\chi_{\rm BH}$ in the parameter space. This is partly due to the fact that the NR waveforms available for high $q$ or high $\chi_{\rm BH}$ are usually shorter compared to the low mass, low spin waveforms. 
\begin{figure*}
\begin{center}
$\begin{array}{cc}
\includegraphics[width=3.in]{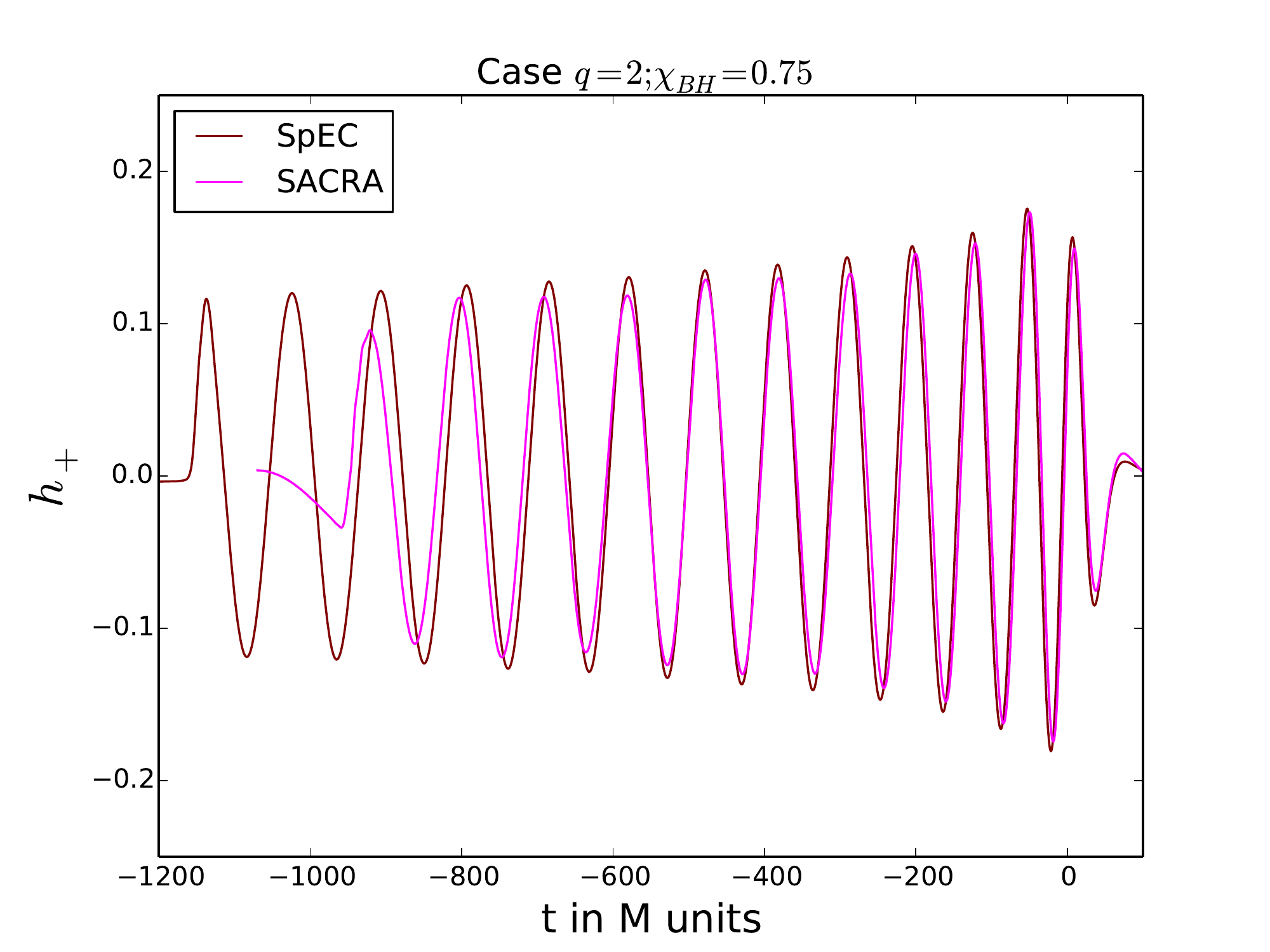}&
\includegraphics[width=3.in]{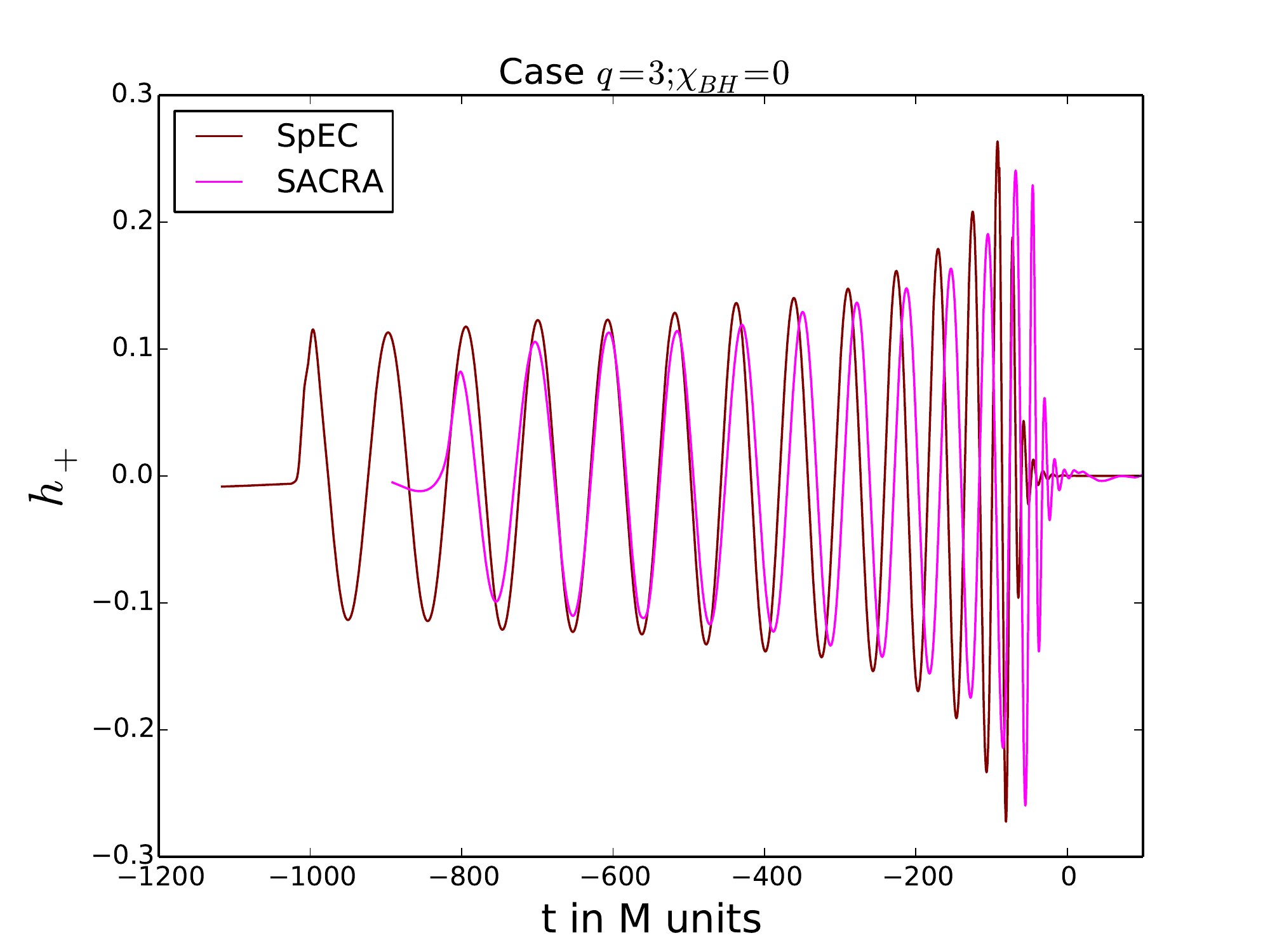}\\
\includegraphics[width=3.in]{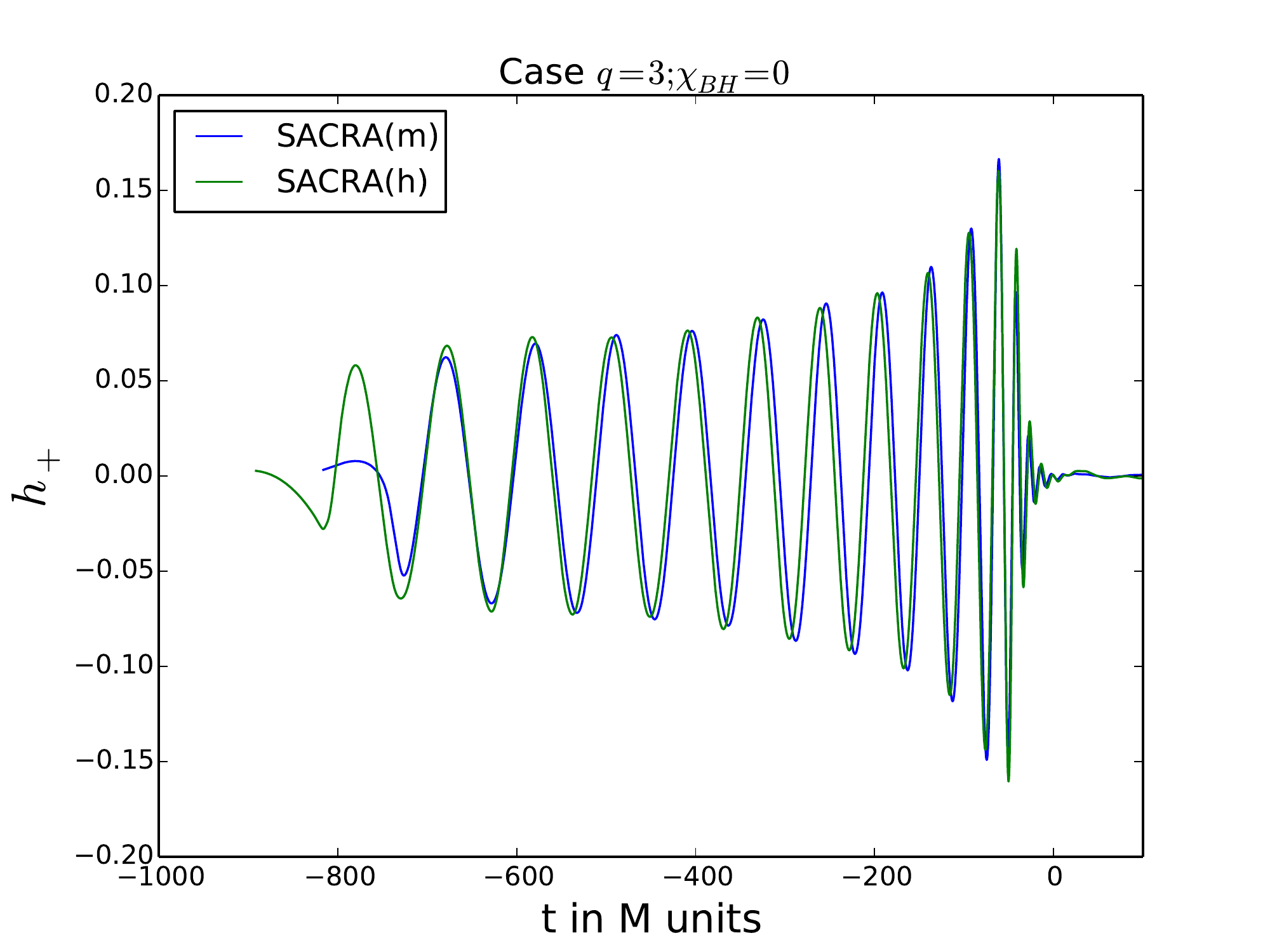}&
\includegraphics[width=3.in]{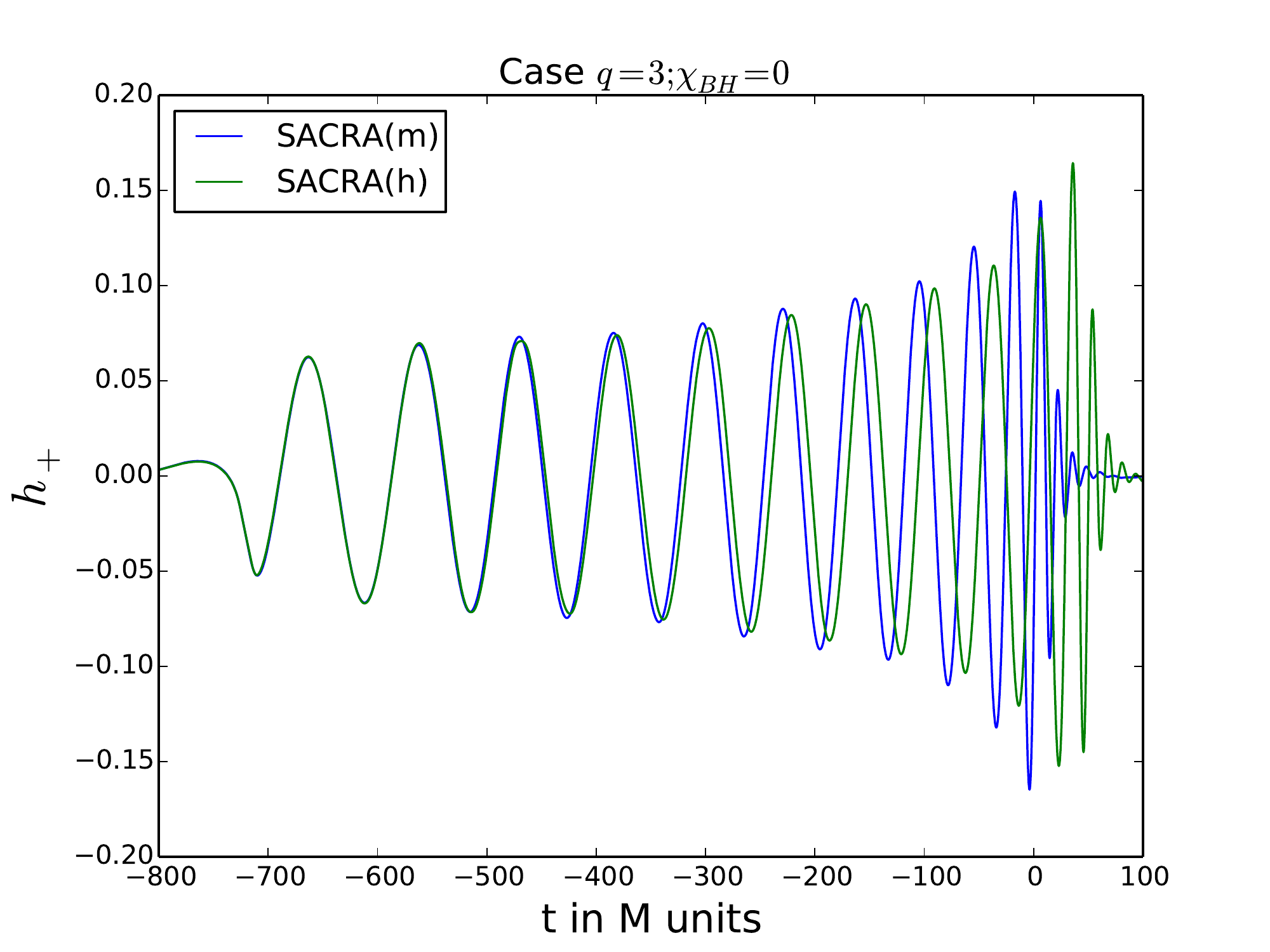}\\
\end{array}$
\end{center}
\caption{In each of the two cases in the top panel, we show the relative phasing between SpEC and SACRA NR cycles when their overlap has already been maximized over initial phase and time for the interval where both types of cycles are present. The case on the left shows good overlap $\sim 95\%$, while that on the right gives poorer overlap $\sim 65\%$. In the left figure of the bottom panel the same maximization is done for two different resolutions of SACRA NR cycles. The resulting overlap between the different resolutions is seen to be $\sim 97\%$. In the right figure of that panel no maximization is done but two different resolutions of SACRA NR cycles are shown with their instantaneous starting frequency aligned. 
}
\label{Fig_Bias}
%
\end{figure*}

\section{Results}
\label{results}

Our goal here is to study the effect of the various sources of systematic errors listed above on the estimation of the tidal deformability parameter. 
In practice, a Bayesian parameter estimation analysis is employed to infer the parameters of the source from its GW signal \cite{Veitch:2014wba}. However, this method is  computationally expensive and beyond the scope this paper (we will present the results for the Bayesian analysis in a future work). Instead we quote the maximum likelihood estimators of the parameters (i.e., parameter values of the best matched template) 
and the true values of the parameter; their respective differences are the errors in their measurement. Below we  briefly describe how we compute the error in the tidal deformability parameter.


Let $h\left(t,\vec{p}_h\right)$ denote a unit-norm hybrid waveform characterized by parameters $\vec{p}_h = \left\lbrace M_{\rm BH}^h, M_{\rm NS}^h, \chi_{\rm BH}^h, \Lambda^h\right\rbrace $. Let $u\left(t, \vec{p}\right)$ be the generic form of a unit-norm template, where $\vec{p}=\left\lbrace M_{\rm BH}, M_{\rm NS}, \chi_{\rm BH}, \Lambda\right\rbrace$ are the dynamical parameters of the binary template.
The inner product between $u$ and $h$, i.e., $\left\langle u\left(f, \vec{p}\right)| {h}\left(f,\vec{p}_h\right) \right\rangle$, maximized over the initial phase and time-of-arrival, is the match. 
The inner product itself is defined for vectors $a$ and $b$ as,
\begin{equation}
\langle a| b \rangle = 4\, {\mathcal Re} \left[\int_{f_{\rm low}}^{f_{\rm high}} \frac{\tilde a(f)\, \tilde b^{*}(f)}{S_h(f)}\, df \right]\,,
\end{equation}
where $S_h(f)$ is the one-sided power spectral density (PSD) of the noise of the detector.  
The fitting factor (FF) is the match maximized over the template parameters:
\begin{equation}
{\rm FF} = \underset {\vec{p},~t_0, \phi_0}{\rm max}~ \left\langle \tilde{u}\left(f, \vec{p} \right)| \tilde{h}\left(f,\vec{p}_h\right) \right\rangle \,.
\end{equation}
The maximization required for computing the FF was carried out using the well known Nelder-Mead downhill simplex algorithm~\cite{Nelder:1965zz}.

\begin{figure*}
\begin{center}
$\begin{array}{cc}
\includegraphics[width=3.in]{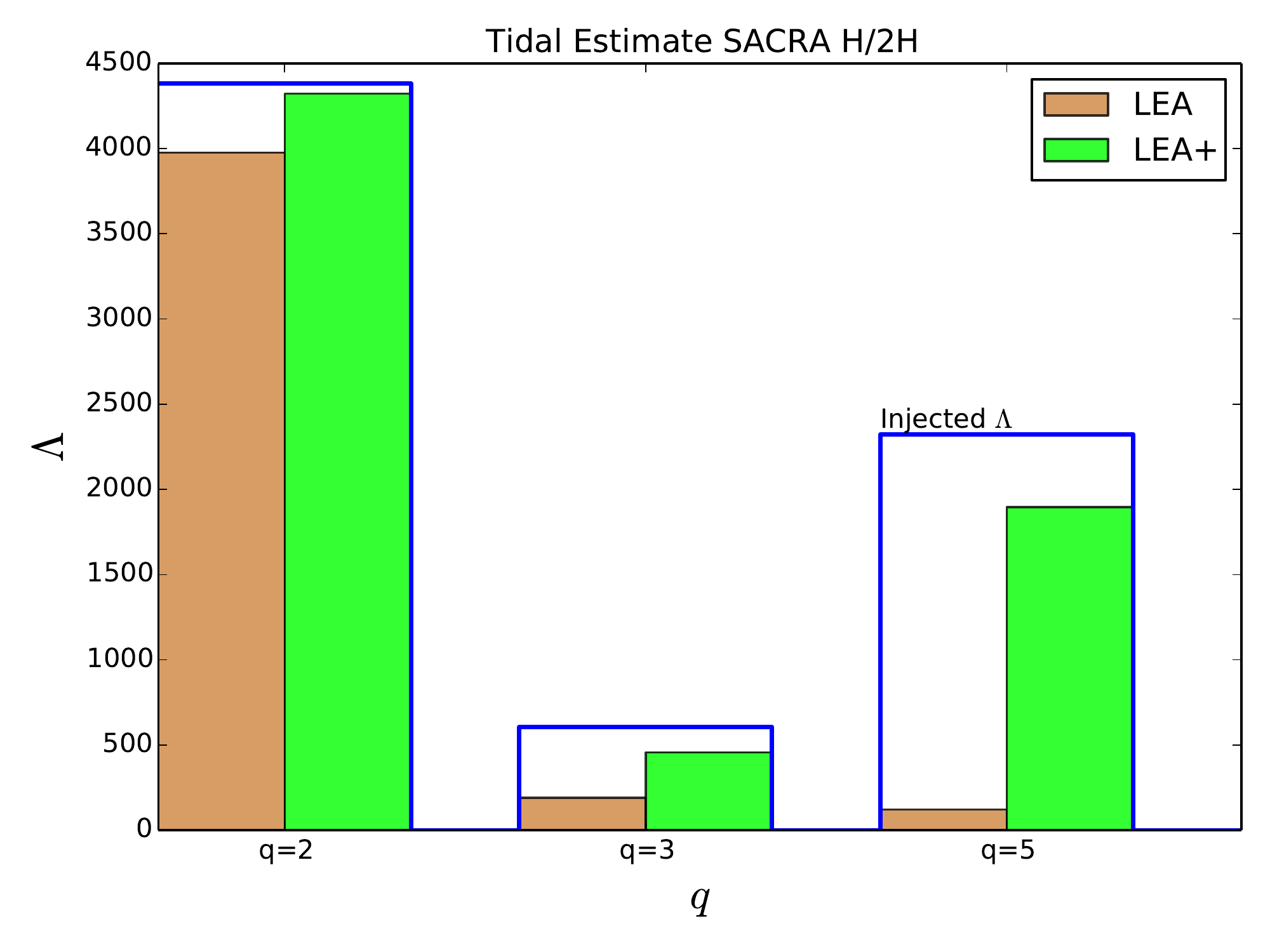}&
\includegraphics[width=3.in]{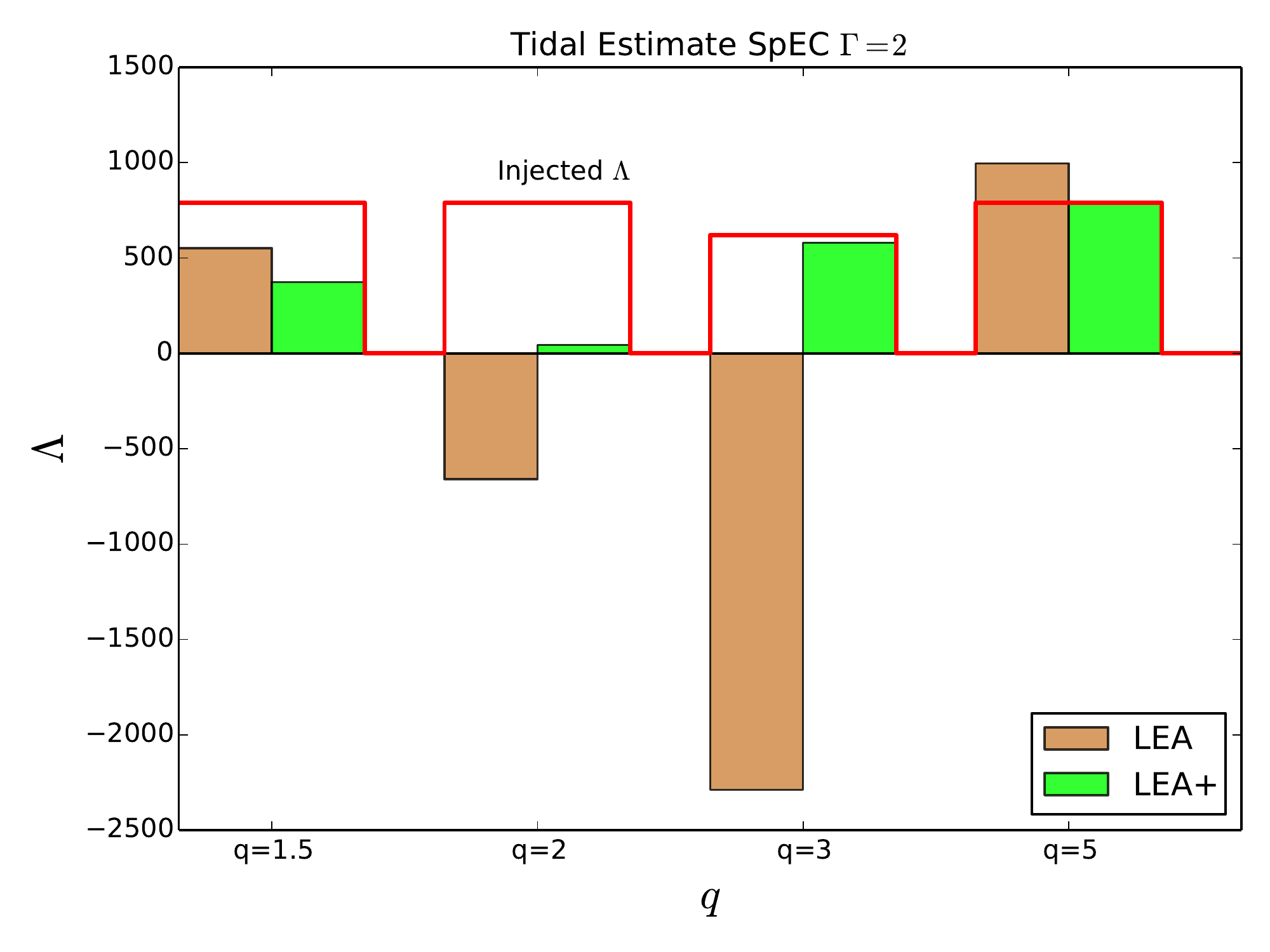}\\
\includegraphics[width=3.in]{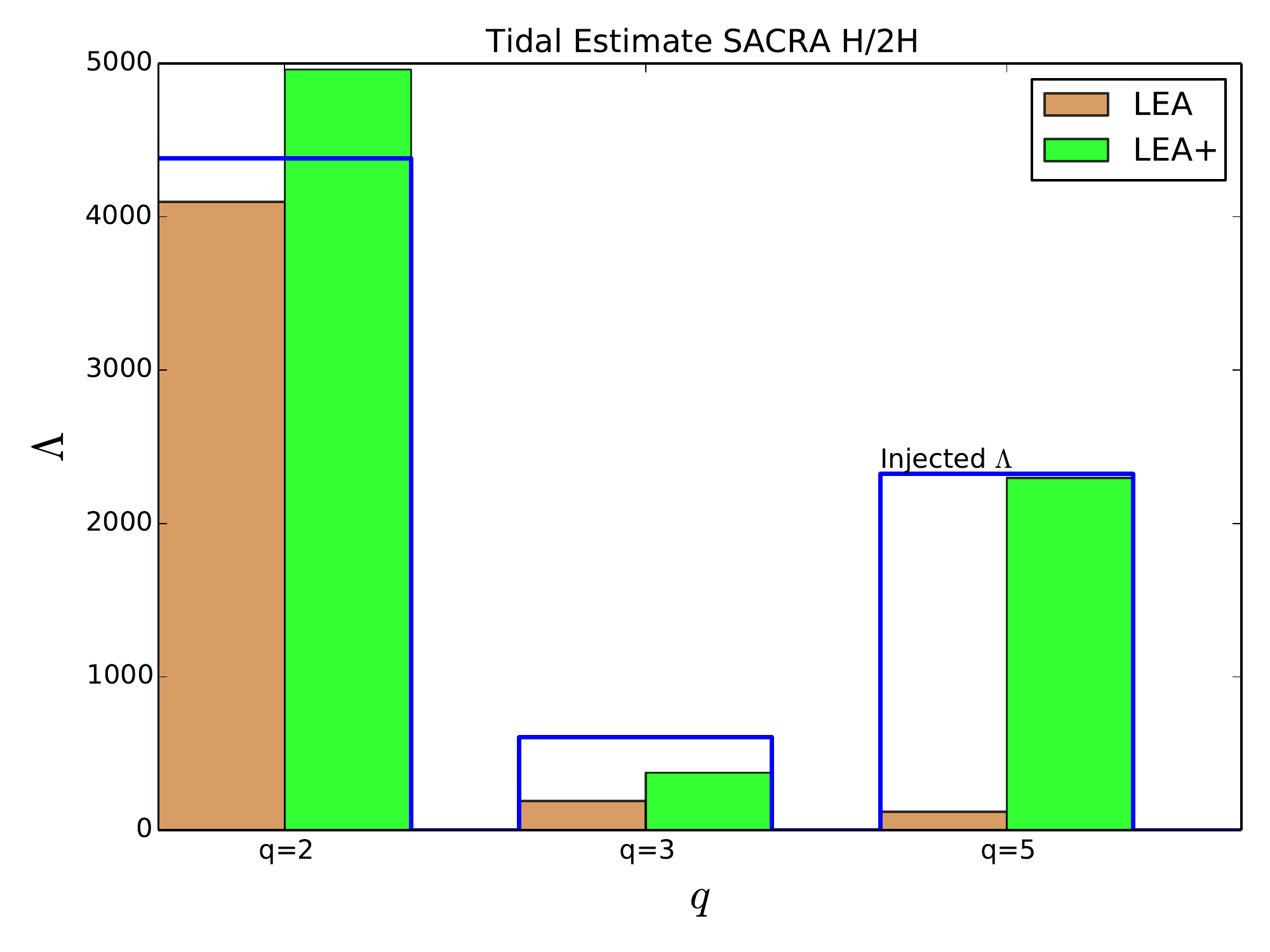}&
\includegraphics[width=3.in]{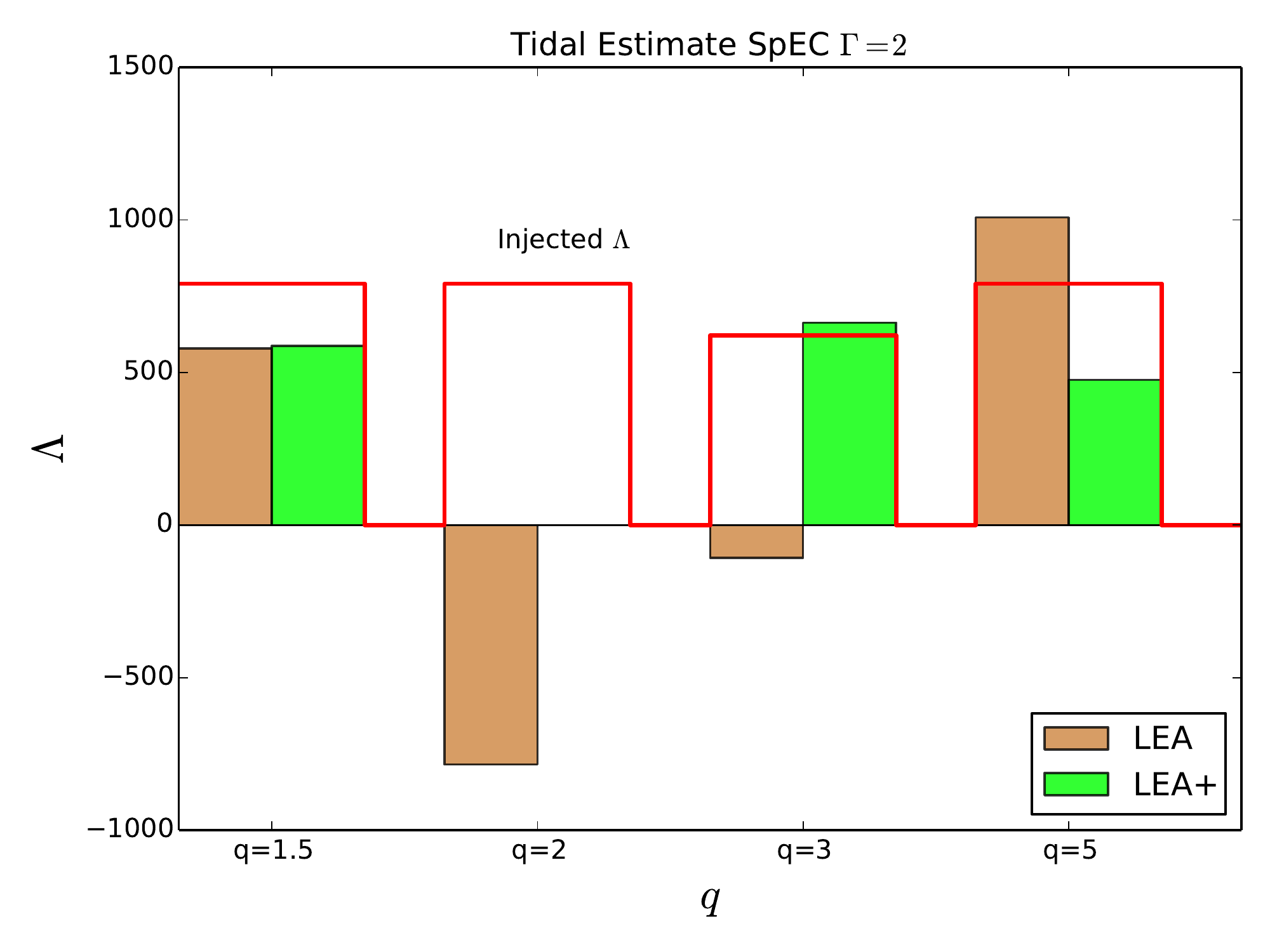}\\
\end{array}$
\end{center}
\caption{Estimates of $\Lambda$ for the highest resolution (top row) and lowest resolution (bottom row) SpEC and SACRA waveforms for the cases listed in Tables~\ref{tab-resolution} and \ref{tab-gesolution}. The red and the blue thick lines denote the true (or injected) value of $\Lambda$ for those two cases, respectively. The brown and green colored bars show the estimated values of $\Lambda$ while employing LEA and LEA+ templates, respectively.}
\label{Fig:pe_lambda}
\end{figure*}

In this paper, we use the 
Zero-Detuned-High-Power (ZDHP) noise-curve of aLIGO \cite{aLIGO_psd_ref} for PSD calculation, and choose $f_{\rm low}$ to be $10$ Hz.
The systematic effects will therefore influence the bias $\Delta p^\mu = (p_0^\mu - p_h^\mu) / p_h^\mu$, where $\mu$ is a parameter index and $p_0^\mu$ denotes the maximizing values of the template parameters.
We study below how large these biases are and explain the reason behind them.


The estimates of the tidal deformability parameter across different NR codes and numerical resolutions are summarized in Tables~\ref{tab-resolution} and \ref{tab-gesolution}. The cases where the step hybridization was applied are indicated in the same tables by asterisks.  Further, estimates of the key non-tidal parameters are presented in Fig.~\ref{Fig:pe}. The faithfulness of different tidal templates (namely, LEA {\it vs} LEA+) in estimating $\Lambda$ can be inferred by comparing Tables~\ref{tab-resolution} and ~\ref{tab-gesolution}.  Using these data, we will estimate the importance of the various systematic errors in turn.

\subsection{Numerical Sensitivity: Comparison of NR codes}

\begin{figure*}[h]
\begin{center}
$\begin{array}{cc}
\includegraphics[width=3.in]{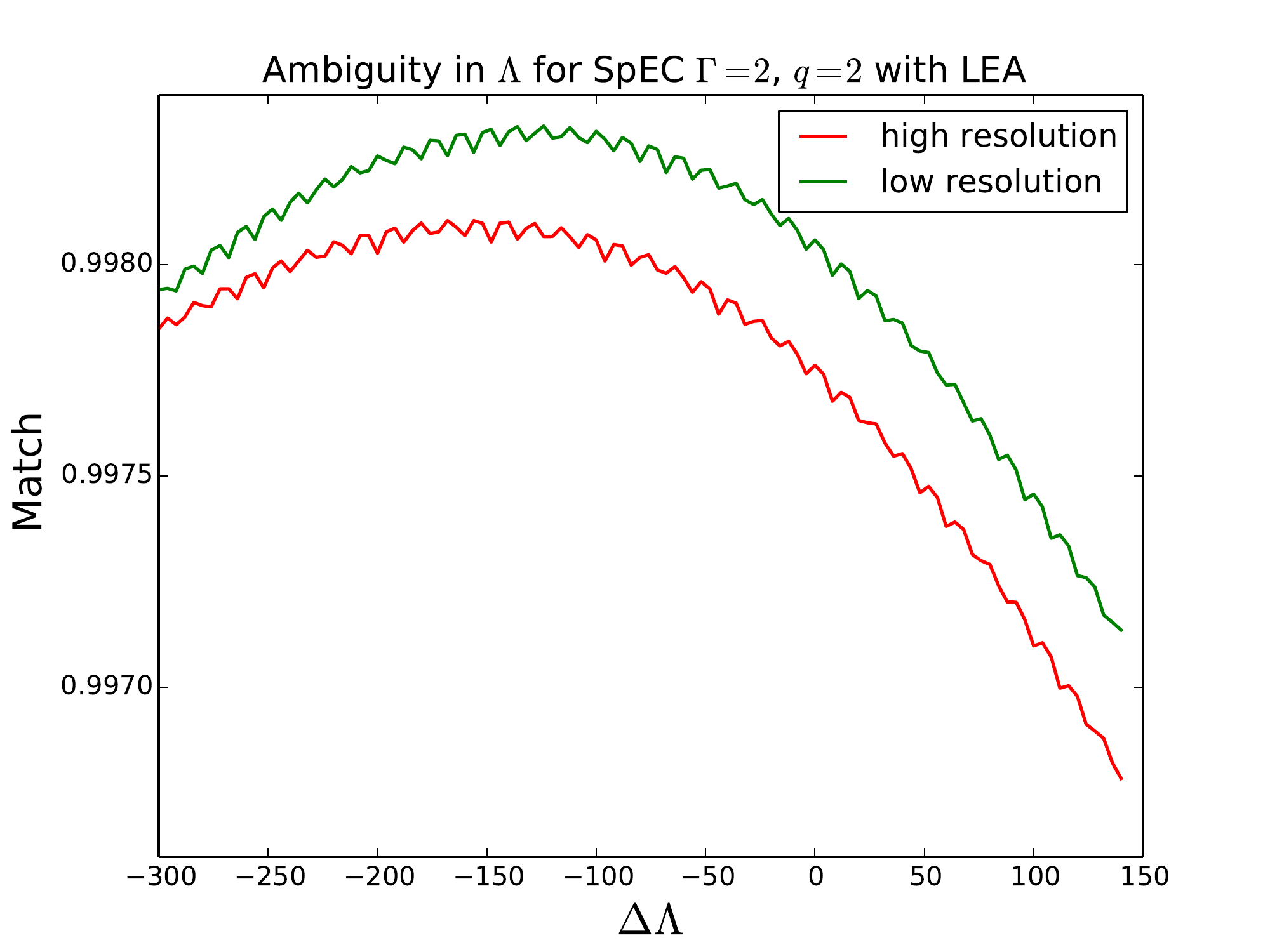}&
\includegraphics[width=3.in]{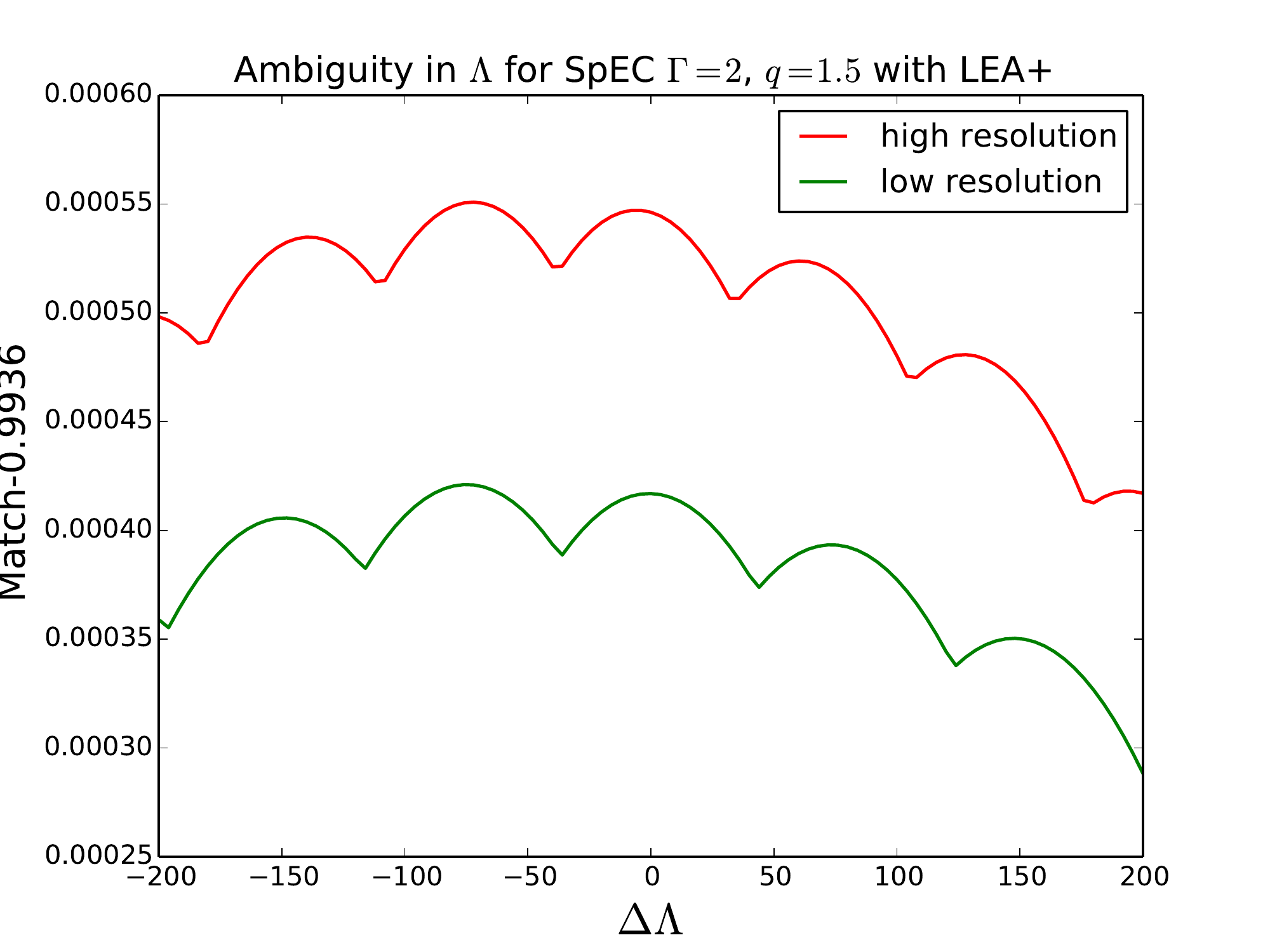}\\
\includegraphics[width=3.in]{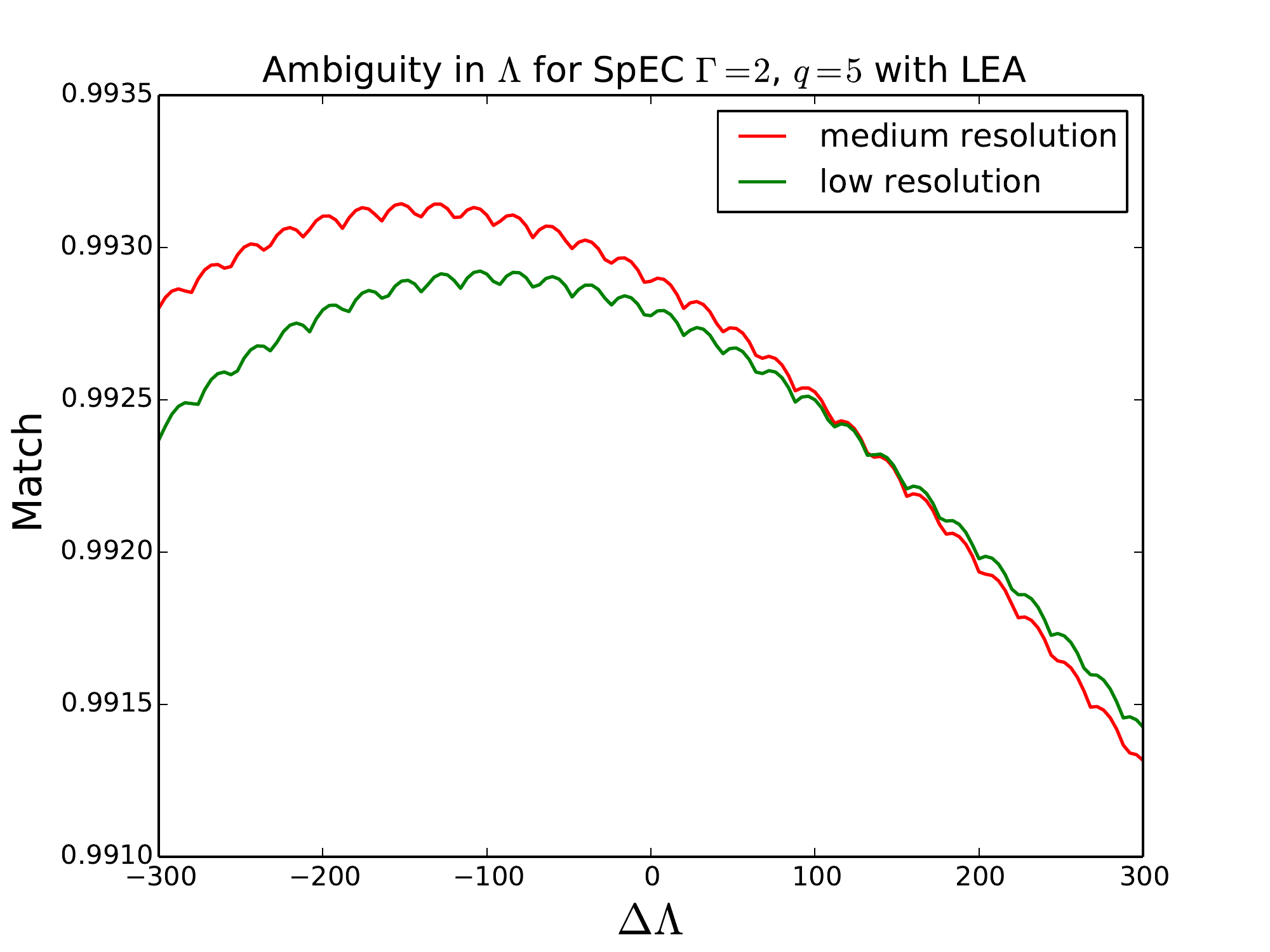}&
\includegraphics[width=3.in]{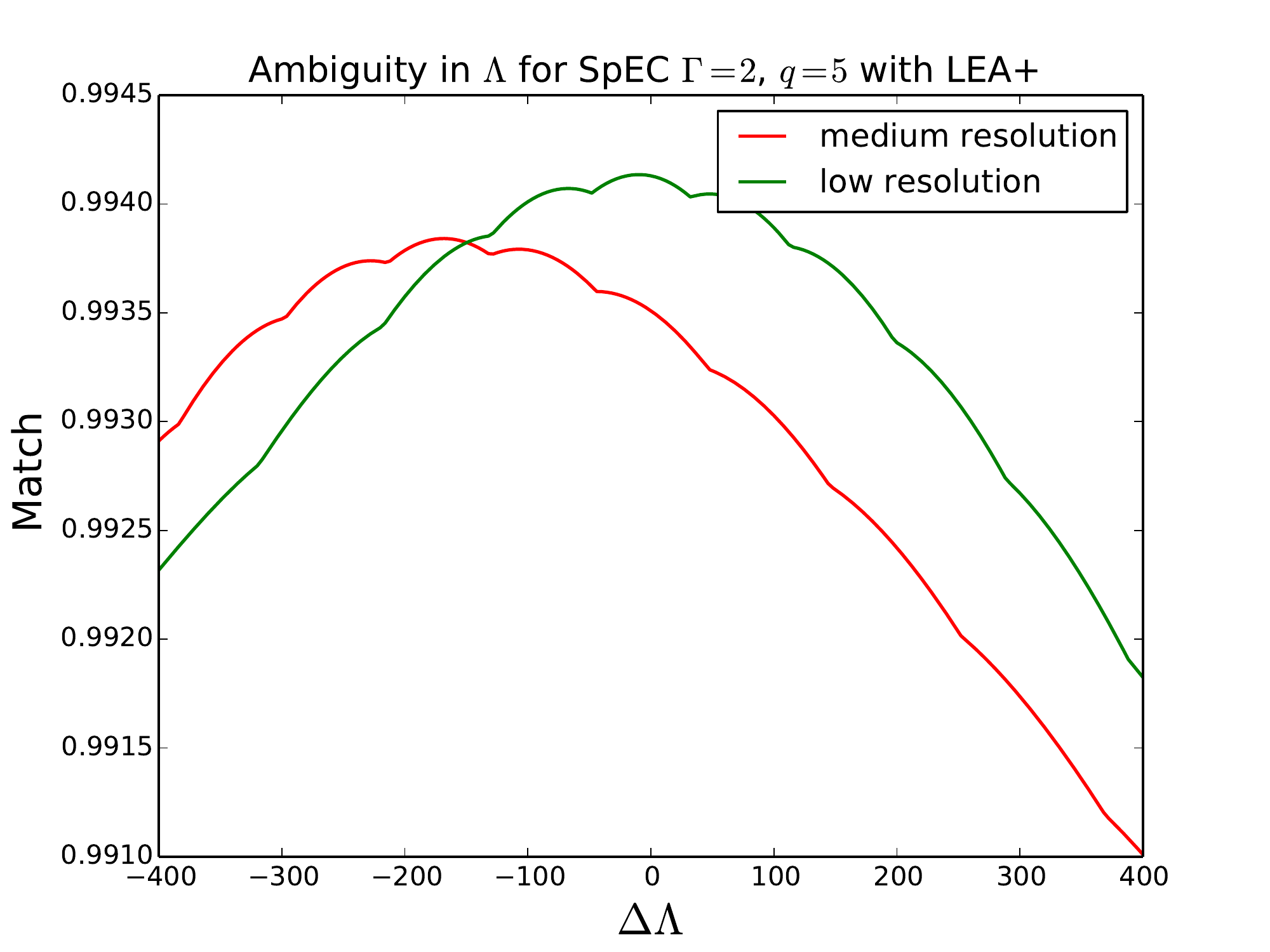}\\
\end{array}$
\end{center}
\caption{Constrained ambiguity functions in $\Lambda$ produced for SpEC hybrids matched with LEA (left panel) and LEA+ (right panel) templates for two different mass ratios, namely, $q=1.5$ and $q=5$. Note that $\Delta \Lambda \equiv \Lambda - \Lambda_{\rm FF}$, where $\Lambda_{\rm FF}$ is the value for which the match is maximum, and is given in Tables~\ref{tab-resolution} and \ref{tab-gesolution}. The sampling rate for computing the above match was chosen to be 4096 Hz. For rates as high as 32 kHz, the small oscillations go away. The concomitant local maxima prevent our fitting factor code from finding the global maximum. Let us illustrate this behavior with the plots in the top panel: The peak in the left figure (high resolution) is at $\Delta \Lambda 
\approx -130$, where $\Lambda \approx 551.3 -130 = 421.3$ is what the FF code should have ideally recovered instead of 551.3 if it had not got stuck at a local maximum (see Table~\ref{tab-resolution}). That would imply a bigger error in $\Lambda$ estimate of about 47\%. (Note that  error covariances of $\Lambda$ with other NSBH parameters can change this estimate somewhat.) The peak in the top right figure (high 
resolution) is at $\Delta \Lambda \approx -80$, where $\Lambda = 373.5 -80 = 293.5$ is what the FF code should have ideally recovered (see Table~\ref{tab-gesolution}), once again, ignoring covariances with other NSBH parameters. That would still imply an error in $\Lambda$ estimate of about 63\%. In all of the cases we observe that $|\Delta \Lambda| \lesssim 200$.}
\label{Fig:ambiguity}
\end{figure*}

\begin{figure*}
\begin{center}
$\begin{array}{cc}
\includegraphics[width=3.in]{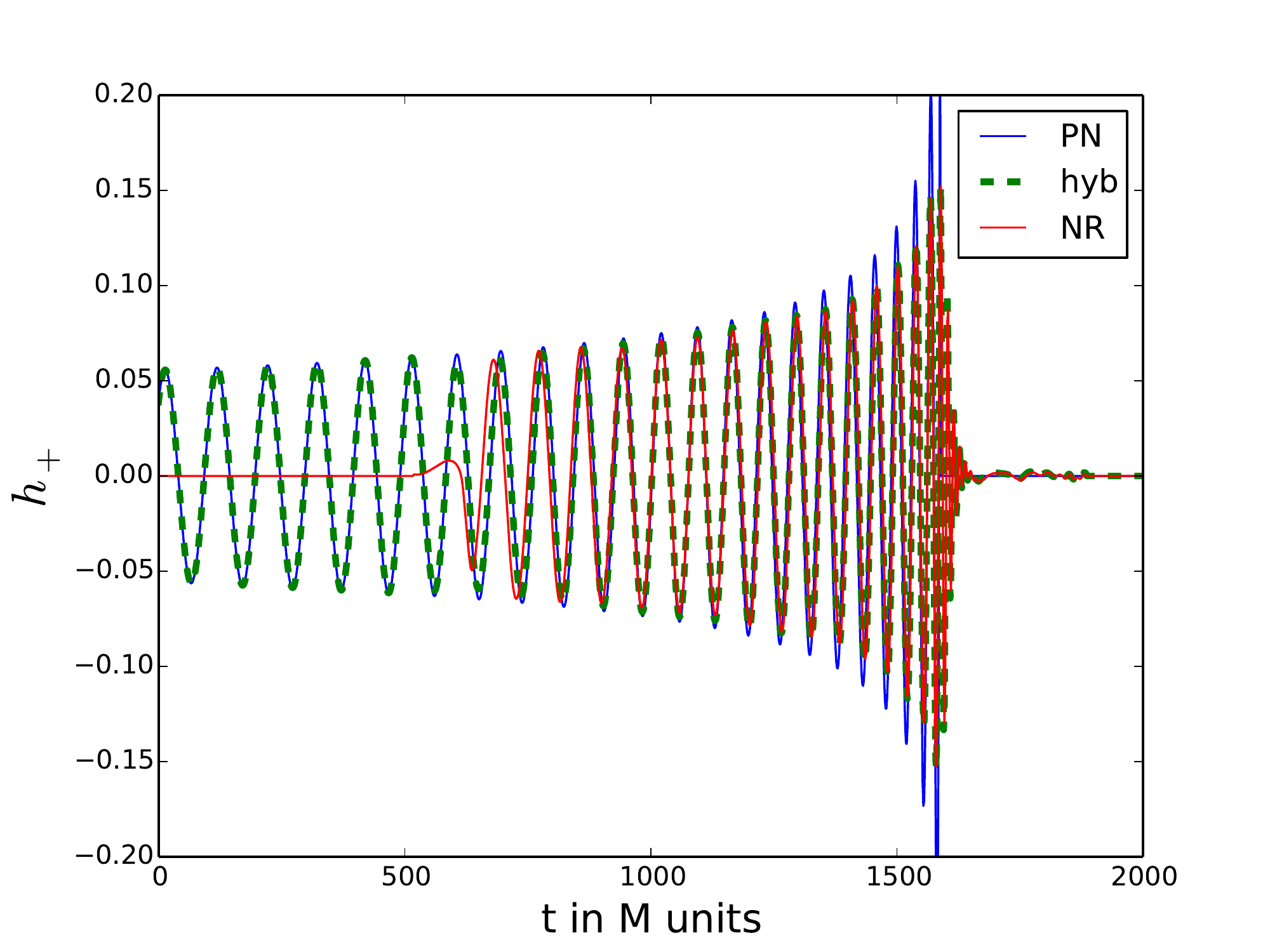}&
\includegraphics[width=3.in]{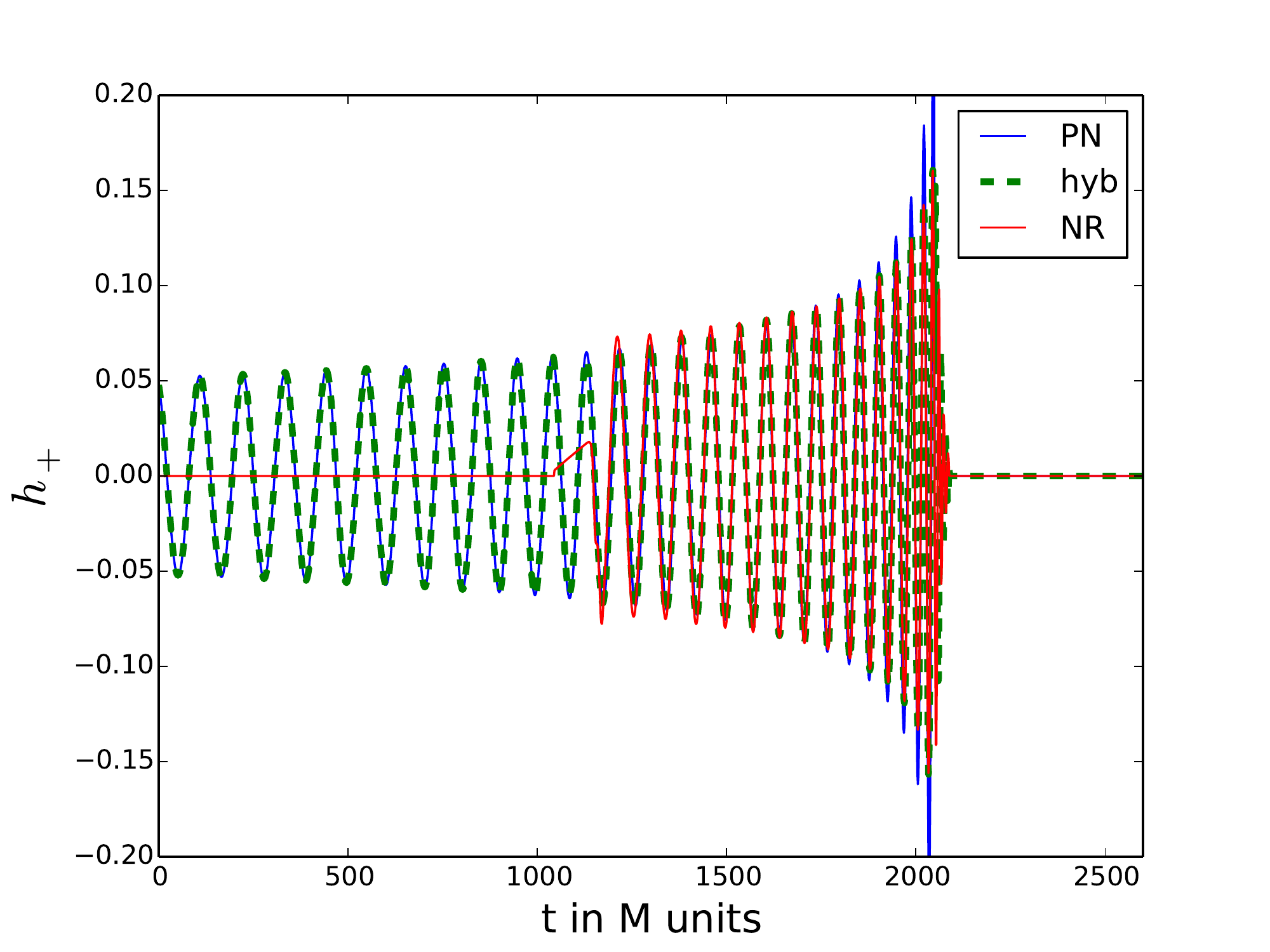}\\

\end{array}$
\end{center}
\caption{Examples of hybrid waveforms generated for SpEC (right) and SACRA (left). Both have $q=5, \chi_{\rm BH}=0.5$   
The time in total mass $M$ at which the instantaneous frequency of the binaries crosses 10 Hz is, respectively, $-7.143 \times 10^6$ and $-7.863\times 10^6$ for SpEC and SACRA. 
}
\label{Fig:hybrids}
\end{figure*}

Since SACRA and SpEC differ in gridding, formulation of Einstein's equations (BSSN~\cite{ShibataNakamura95,BaumgarteShapiro98} {\it vs} generalized harmonic~\cite{Lindblom:2007}), method of treating black holes (moving punctures {\it vs} 
excision), and gauge, 
that the two codes give consistent answers for gauge-invariant outputs, up to the truncation errors of each code, is an important test of both codes.  For this test, we pick two binary configurations used in the construction of LEA and simulate them with the SpEC code. For both cases and for both codes, only the last roughly 5 orbits prior to merger are simulated.  The initial data of the SpEC simulations has undergone an eccentricity reduction procedure, so the initial eccentricity for these runs (0.005--0.008) may be slightly lower than that of the SACRA runs.

For our first comparison, we choose an extreme case that maximizes the importance of tidal effects:  low mass-ratio ($q=2$), high BH spin ($\chi_{\rm BH}=0.75$), and a large (even if somewhat unnatural in the light of GW170817 \cite{GW170817}), low-mass NS ($M_{\rm NS}=1.2 M_{\odot}$, $\Lambda=4382$). This large NS radius is a consequence of the very stiff piecewise-polytropic EOS used in this case, labeled ``2H'' in~\cite{Lackey_2014}. 
Tidal effects are easy for both codes to resolve for this case. The results are listed in Case 3 in Table ~\ref{tab-resolution}, and the error in $\Lambda$ turns out to be less than $10\%$ for both codes.

For the second comparison, we choose a binary system close to the center of the LEA parameter space, namely, with $q=3$, $\chi_{\rm BH}=0$, $M_{\rm NS}=1.35 M_{\odot}$, $\Lambda=607$.  For this configuration, listed as Case 5 in Table~\ref{tab-resolution}, the tidal effects are difficult to resolve, but we have as many as 3 SACRA resolutions and 2 SpEC resolutions to estimate errors. 
With the LEA templates, differences between the recovered $\Lambda$ for different numerical resolutions is up to ten times the injected value of $\Lambda$, and we are thus entirely unable to reliably recover $\Lambda$. With LEA+, the recovered $\Lambda$ is order-of-magnitude accurate, despite larger errors in the masses of the compact objects.


Next, we compare the $l=2, m=2$ mode of the decomposition of numerical waveforms in spherical harmonics in Fig.~\ref{Fig_Bias}.
In the top panel of that figure, we show for two different binaries the SACRA waveform, time-shifted relative to the SpEC results to maximize the match between waveforms. The top left panel plots SACRA and SpEC waveforms for the first (high spin, 2H EOS) case.  For this ``easy'' case, we find reasonable agreement, with a match around 90\%.  However, the agreement between SpEC and SACRA is not as good for the second (nonspinning, H EOS) case, shown on the top right of Fig.~\ref{Fig_Bias}.  For this case, visible dephasing is seen, and the match is lower than 70\%. This disagreement indicates a large error in at least one set of simulations.  It is only a sign of inconsistency between the two codes if it is larger than the numerical errors of the individual simulations.  We thus turn to a consideration of the error of each code's simulations, revealed by differences between resolutions.

A second SpEC simulation at higher resolution agrees with the lower-resolution to 0.1 radian through inspiral and 0.5 radian at the end of merger. However, details of SpEC's adaptive mesh refinement algorithm can occasionally lead to significant overestimates of the numerical errors when only 2 simulations are used to estimate these errors. While the numerical errors are consistent with those measured in other recent SpEC NSBH simulations, a more rigorous error estimate would require a third simulation. On the other hand, the two available resolutions allow us at least to test the effect on parameter estimation of this level of phase error. 

Convergence of the SACRA run is illustrated in the bottom panels of Fig.~\ref{Fig_Bias}.  Since these simulations are several years old, the accuracy is lower than current SACRA simulations.  If the time is shifted for optimal match, the agreement of different SACRA resolutions is reasonably good. For purposes of detection, this is the most important convergence check.  If time shifting is not done (since different resolutions start from the same initial state), dephasing comparable to the SpEC-SACRA difference is seen.  This more pessimistic comparison is more relevant for parameter estimation errors.  That is, the ability to optimize match to a high value may give overly optimistic expectations for parameter estimation errors.
A longer waveform would also make the first comparison a lot worse than it is - the only reason this time shift works is because the numerical error happens to roughly compensate the change in inspiral rate at a different separation for a large portion of the very short simulation.

\subsection{Numerical Sensitivity: Effect of Resolution and the base BBH waveforms used}

\subsubsection{Effect of resolution on parameter estimates}

One way to test the importance of numerical truncation error is to find the best template matches to numerical waveforms generated from simulations of different resolutions.  We compare best matched parameters for each binary system for which we have multiple numerical resolutions.  The results are listed in Table~\ref{tab-resolution} for best matches to LEA templates and Table~\ref{tab-gesolution} for best matches to LEA+ templates.

Several conclusions are apparent.  First, the $\Lambda$ estimates are indeed often sensitive to the NR contribution, as can be seen from the effect of altering resolution. One might imagine that the many PN cycles with the correct tidal contributions would ``override'' the effect of numerical error in the last few cycles, but this does not turn out to be the case, in general. Including correct PN tidal terms in the inspiral part of the waveform will not guarantee that the $\Lambda$ estimate remains unaltered with changing resolution. Second, tidal effects are seen to be recovered with somewhat small errors for the low mass case $q=1.5$ in LEA. 
Tidal effects are also seen to be recovered accurately for high spin systems, for which the NS disrupts well outside the innermost stable circular orbit even for moderate mass ratios, leaving a strong, easily resolved imprint in the waveform.  On the other hand, for $q\ge 3$ and low spin the deviations between resolutions, even for resolutions with lowest phase difference, 
swamp the measurement of $\Lambda$.  We see that in many cases, the best-fit template does not even have positive $\Lambda$.  For $q=2$, $\chi_{\rm BH}=0$, numerical accuracy does appear to be sufficient, in that recovered parameters change little with resolution.  But here there are other sources of systematic error which again swamp the physical tidal effect and produce negative recovered $\Lambda$. 

Why does the estimated $\Lambda$ often vary so widely with resolution?  One interpretation would be that the numerical truncation error is swamping the physical tidal effect.  However, there is another possibility. If a template family has multiple members with match close to the maximum, small differences between resolutions might cause large jumps in parameter estimates.  We explore this possibility below in the following section.

\subsubsection{Effect of the base BBH waveforms}

We performed a detailed comparison of the estimates of tidal and non-tidal parameters by LEA and LEA+ (see Tables~\ref{tab-resolution} and \ref{tab-gesolution}). We will mainly focus on the overall trend in the estimation of the tidal parameter $\Lambda$ (see Fig.~\ref{Fig:pe_lambda} as well). It is immediately clear that LEA+ does better at returning physically acceptable (non-negative) values of $\Lambda$. Given that LEA and LEA+ have 
similar tidal terms in their waveform models, we reason that the choice of the base BBH models, namely IMRPhenomC and
SEOBNRv2, respectively, is likely responsible for most of these differences.

A comparison of Tables~\ref{tab-resolution} and \ref{tab-gesolution} indicates that the sensitivity of $\Lambda$ estimates to numerical resolution is itself PN template-dependent.  Notice that the variation in $\Lambda$ with resolution is much larger for Case 1 when LEA+ templates are used than when LEA templates are used.  For Case 5, the opposite is seen:  good consistency in $\Lambda$ for estimation using LEA+, but wide variation when LEA is used.  From the behavior of Case 1 in LEA+, we see that parameters differ most at the highest resolution, a result that seems to contradict the good convergent behavior seen in the numerical waveforms. Extreme sensitivity to the waveform (or the lack thereof) can result from the shape of the match as a function of the deviation in the values of the template parameters from those of the signal.  

Indeed, estimating the value of a parameter then involves maximizing the match of normalized templates with the data containing the signal.  In Fig.~\ref{Fig:ambiguity} we show that the rate at which the match changes with a deviation in the value of $\Lambda$ from that of the signal is slow in the sense that the match drops by a few tenths of a percent even when that deviation is as high as several tens to a few hundreds. This is because the ambiguity function~\cite{Dhurandhar:2017aan} in $\Lambda$ is diffuse\footnote{In this case, the ambiguity function is the match between two unit-norm templates, from the same waveform family, with different values of $\Lambda$.}. What is plotted, however, is not the ambiguity function itself since the hybrid waveform and the template are {\rm not} from the same waveform family. We will call this quantity the {\em constrained} ambiguity function since the templates, in general, may be constrained to reside in a subspace of the data-space that does not fully overlap with the subspace in which the hybrid waveform resides. Additionally, the same figure shows that this function has multiple local maxima where a parameter estimation algorithm can get stuck and miss finding the global maximum. This effect introduces a fraction of the error in $\Lambda$ estimation. 
The small oscillations, accompanied by the local maxima and minima, in the constrained ambiguity function arise when the sampling rate of the data and the templates is not high enough (Fig.~\ref{Fig:ambiguity} used 4096 Hz) and 
has been studied in a somewhat different context in Ref.~\cite{Ajith:2012az}. As illustrated in Fig.~\ref{Fig:ambiguity} this effect can contribute to the systematic error in the estimation of $\Lambda$, and its extent is somewhat different for LEA and LEA+. 

\begin{figure*}
\begin{center}
$\begin{array}{cc}
\includegraphics[width=3.in]{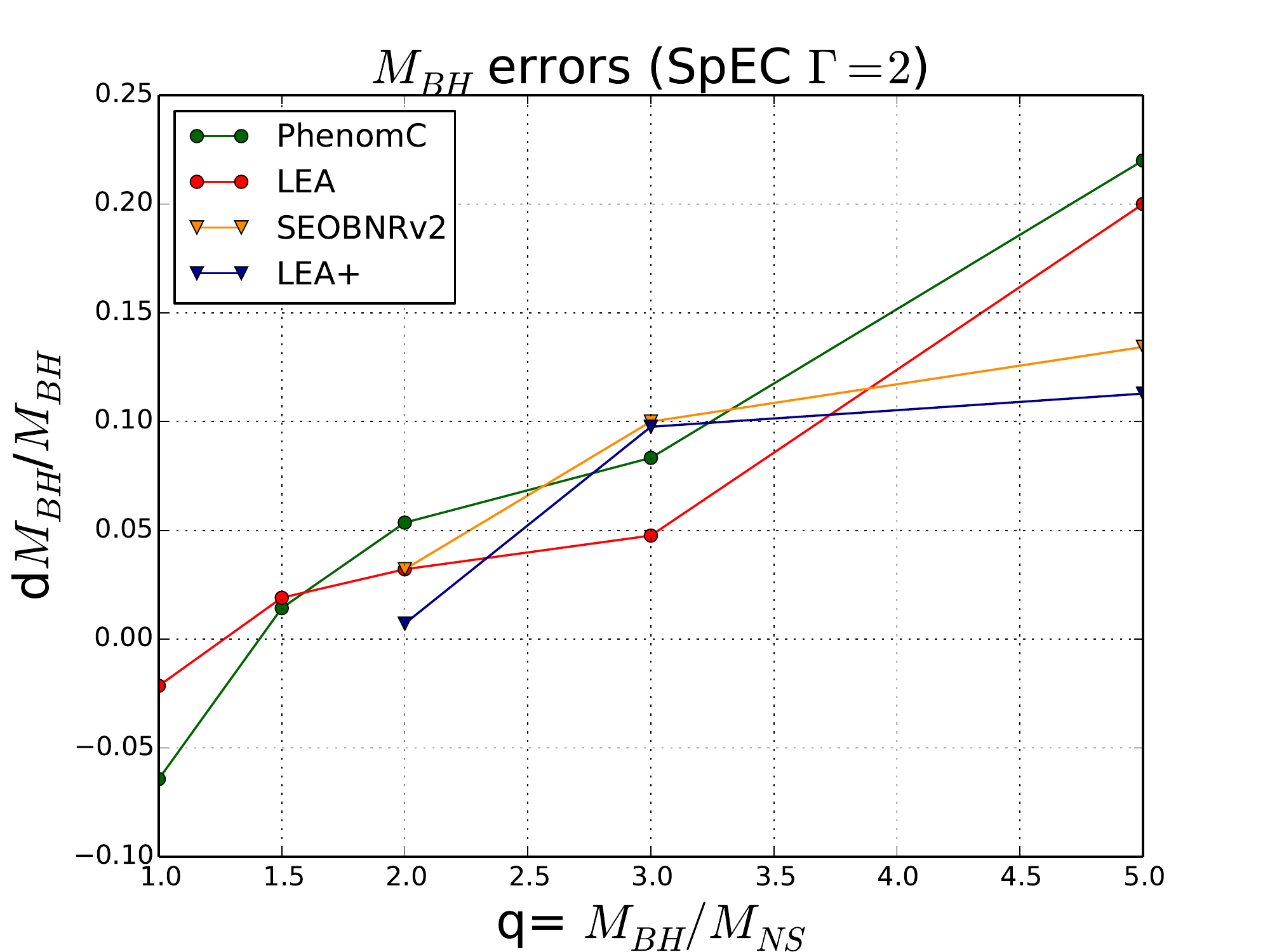}&
\includegraphics[width=3.in]{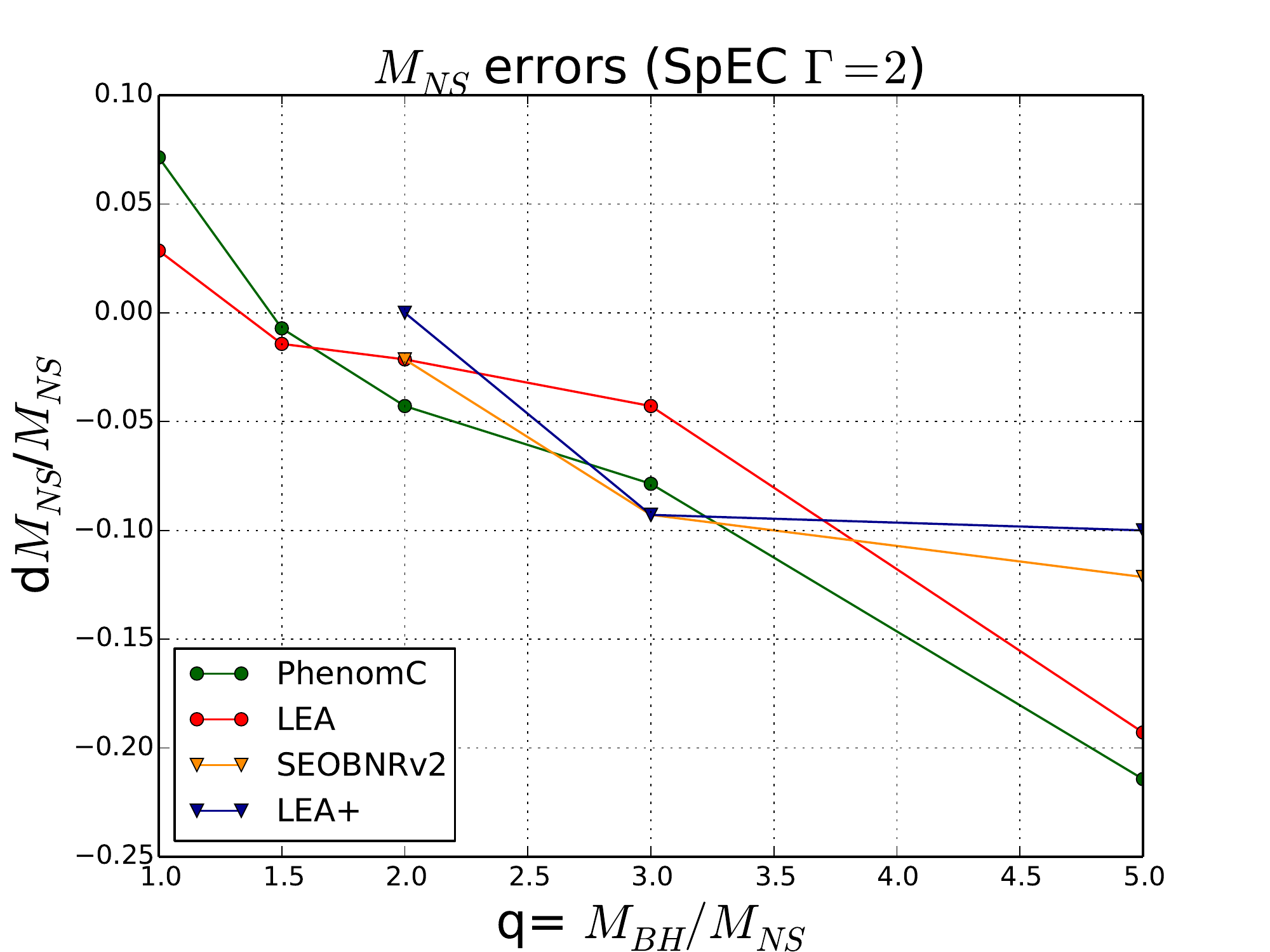}\\
\end{array}$
\end{center}
\caption{Component mass estimation errors for templates based on two PN (IMRPhenomC \& LEA) and two EOB (SEOBNRv2 \& LEA+) models. Apart from $q=5$, which has $\chi_{\rm BH}=0.5$, all other systems shown above are non-spinning.}
\label{Fig:pe}
\end{figure*}

\subsection{Challenges for hybridization}

As we noted before, the regular technique of hybridization is likely to fail in the presence of high $q$ or $\chi_{\rm BH}$. For such cases, we have no other option but to rely on the step method. While theoretically inferior to
the regular method, the step method
yields positive tidal estimates in all of the waveforms where this procedure is employed. This is possibly because this latter procedure emphasizes
continuity of phasing frequency between PN and NR cycles, sometimes at the cost of a better match. All cases in point have been highlighted in asterisk in Tables~\ref{tab-resolution} and \ref{tab-gesolution}. We do not claim this to be a better method in general, but a useful one
when the standard method fails. 

These hybridization issues are affected by
the number of NR cycles utilized by that procedure. Pursuing the simulation of sufficiently long 
NR waveforms is a worthwhile goal for achieving reliable hybrid waveforms. With such NR input, step hybridization will not be necessary at all, because all the waveforms will be guaranteed a relatively large patching region. These waveforms will be critical for producing more accurate NSBH templates, especially, for estimating $\Lambda$. 

In the same way, having fewer NR cycles 
implies a smaller patching region thus making hybridization by method of Eqs.~(\ref{eq:delta}) and (\ref{eq:hybrid}) more likely to fail. Once a stable patching region is established involving a sufficiently large number of cycles from accurate overlapping regions of PN and NR waveforms, the tidal estimate is expected to vary minimally with addition of more NR cycles -- a fact that has been demonstrated conclusively in~\cite{MacDonald:2011ne}. It is also known from the same work that for some of the non-tidal parameters their accuracy depends on where in the frequency domain the overlapping region is chosen for hybridizing the waveforms. The important message here is this: given an SNR value, there exists an upper bound on the number of NR cycles necessary for the construction of a hybrid waveform of a desired accuracy. Going above this bound will make the systematic error arising from hybridization subdominant to the statistical error.

Finally, for the few representative cases where both the regular and step methods worked, the estimates of $\Lambda$ are seen to agree to within $1\%$ of the estimate of the unhybridised waveform. This provides a measure of the inherent error that our hybridization method can by itself introduce in the complete waveforms.

\subsection{Binary parameter estimates}

Before concluding, we state the results of the estimates for the non-tidal parameters, namely, $M_{\rm BH}, M_{\rm NS}$ and $\chi_{\rm BH}$. The results of the parameter estimation of masses of the $\Gamma=2$ SpEC waveforms against different tidal as well as non-tidal templates are shown in Fig.~\ref{Fig:pe}. For estimates of the masses, the templates show consistent trends. Errors in the masses are  small for the low-mass binaries, at a few to several percent, and increase with $q$ to $\lesssim 22\%$ for $q=5$. An important contributor to this error is the unfaithfulness of the base waveforms (IMRPhenomC or SEOBNRv2) relative to the NR waveforms on which they are modeled, which itself can cause a bias in the mass parameter of several percent~\cite{Kumar:2016dhh}. Differences in the base PN waveform and the hybridization procedure can add to this error as well. Nevertheless, for low mass-ratios it is expected that these biases would be comparable, if not subdominant compared to the other sources of error we have highlighted.

In addition to the estimation of the individual masses, we note in Tables~\ref{tab-resolution} and ~\ref{tab-gesolution} that the error in  $\chi_{\rm BH}$ can be as high as 14\%,

\section{concluding remarks}
\label{end}

In this paper, we classified some of the sources of systematic errors in NSBH tidal waveforms and demonstrated their significant effect on  accuracy of NS tidal parameter estimation. Importantly, even though the fitting factor values of the NR-based tidal NSBH templates LEA and LEA+ are very high, their best-matched values of the tidal parameter $\Lambda$ show a bias of tens to several tens of percent relative to the true value. In addition to the magnitude of the systematic error in $\Lambda$, our study allows us to draw several conclusions about the nature of this error.

First, we learn that the final cycles, the portion modeled by numerical relativity, have a significant effect on the systematic error even though they are a small part of the waveform.  The modeling of this portion of the evolution is confirmed to be an important endeavor.

Second, we find that the binary BH base of the template family has a large effect on the measured $\Lambda$.  The IMRPhenomC used by LEA, for example, is insufficiently accurate for mass ratios $q>2$.

Third, use of undersampled data or templates can contribute to systematic error in $\Lambda$, as evidenced in the appearance of multiple local maxima in the constrained ambiguity function shown in Fig.~\ref{Fig:ambiguity}.

Fourth, there are nevertheless significant errors in some NR waveforms in use.  Meaningful consistency between SACRA and SpEC could not be established in all cases.  (Errors were too large to definitely establish inconsistency as well.)  This comparison should be continued with each group's more-accurate current code.  These errors have rather little effect on the match and hence on detectability, but they have a modest effect on parameter estimates.


Finally, the number of cycles in some of the available NR waveforms is inadequate to avoid problems in the hybridization process and that longer NR waveforms can mitigate the greater part of this issue.

One could, in principle, try to circumvent many of the problems by restricting efforts in the inspiral-only regime. But this quick fix suffers from obvious shortcomings, apart from losing valuable SNR, we also lose valuable information on the EOS, because the EOS effects are strongest just before merger. A further point is that the agreement of the PN and the numerical waveforms is another strong sanity check on the overall consistency of the waveforms. Resorting to PN only techniques will also rob us of this check.

Evidently the important lesson drawn from this exercise is that concerted effort among NR groups is needed to develop accurate NR waveforms with at least a few tens of cycles that can be used to construct longer NSBH waveforms, through hybridization or calibration, for reliable tidal parameter estimation. With this in mind, our recommendations for any possible future attempts on NSBH waveform construction are as follows. First, the focus of templates should crucially involve the low to intermediate mass ratio $2\leq q \leq 5$ because of stronger prevalence of tidal effects. Targeting the construction of accurate waveforms with a wide range of BH spin
is also desirable since stellar mass black holes are known to have low to high spins, even though there is no information available yet on how large these spins might be in NSBH systems.
A part of this space of physical parameters was already explored in LEA, but with tidal pieces calibrated solely with SACRA waveforms that were old and low in accuracy. We propose that until more accurate tidal templates are found, LEA+ should be used for tidal parameter estimation in real-data searches. The gold standard is to have templates that are tidally calibrated across multiple families of high-quality numerical waveforms with consistently high accuracy.


\vfill\pagebreak

\section*{Acknowledgments}

We thank Bhooshan Gadre and Archisman Ghosh for helpful discussions. We also thank Prayush Kumar for carefully reading the manuscript and making useful suggestions. KC and SB acknowledge support from the Navajbai Ratan Tata Trust.
AG acknowledges support from SERB-NPDF grant (PDF/2015/000263), NSF grants No. AST-1716394 and No. AST-1708146, and the Charles E. Kaufman Foundation of The Pittsburgh Foundation. FF gratefully acknowledges support from NASA grant 80NSSC18K0565.  KK is supported by Japanese Society for the Promotion of Science (JSPS) Kakenhi Grant-in-Aid for Scientific Research (No.~JP16H06342, No.~JP17H01131, No.~JP18H04595). MD acknowledges support through NSF Grant PHY-1806207.  LK acknowledges support from NSF grant PHY-1606654,
and MS from NSF Grants PHY-1708212, PHY-1708213, and PHY-1404569.
LK and MS also thank the Sherman Fairchild Foundation for their support.  This project has been assigned preprint number LIGO-P1800191.





\end{document}